\documentclass[10pt,twocolumn,a4paper]{IEEEtran}
\usepackage{amsmath}
\usepackage{graphicx}
\usepackage{amsfonts}
\usepackage{amssymb}
\usepackage{amsthm}
\usepackage{color}
\usepackage{float}
\usepackage{verbatim}
\usepackage{geometry}
\usepackage{lineno}
\usepackage{caption}    
\usepackage{subcaption}
\usepackage{placeins}
\usepackage{multicol}

\def\à{\`{a}}
\def\ä{\"{a}}
\def\â{\^{a}}
\def\é{\'{e}}
\def\è{\`{e}}
\def\ê{\^{e}}
\def\ë{\"{e}}
\def\ï{\"{i}}
\def\î{\^{i}}
\def\ö{\"{o}}
\def\ô{\^{o}}
\def\ù{\`{u}}
\def\ü{\"{u}}
\def\û{\^{u}}
\def\ç{\c{c}}

\def\kvec{\vec{k}}

\def\zvec{\vec{z}}
\def\green{\textcolor{green}}

\newcommand{\be}{\begin{equation}}
\newcommand{\en}{\end{equation}}

\newcommand{\bea}{\begin{eqnarray}}
\newcommand{\ena}{\end{eqnarray}}

\newcommand{\beano}{\begin{eqnarray*}}
\newcommand{\enano}{\end{eqnarray*}}

\newcommand{\bei}{\begin{itemize}}
\newcommand{\eni}{\end{itemize}}

\def\green{\textcolor{green}}

\providecommand{\tabularnewline}{\\}
\floatstyle{ruled}
\newfloat{algorithm}{tbp}{loa}
\providecommand{\algorithmname}{Algorithm}
\floatname{algorithm}{\protect\algorithmname}

\setpagewiselinenumbers
\modulolinenumbers[5]
\makeatother

\begin{document}
\title{Prediction of Transformed (DCT) Video Coding Residual for Video Compression}
\author{
Matthieu~Moinard, Isabelle~Amonou, Pierre~Duhamel,~\IEEEmembership{Fellow,~IEEE,}~and~Patrice~Brault,~\IEEEmembership{Senior~Member,~IEEE}.%
\thanks{M. Moinard and I. Amonou are with Orange Labs, Rennes, France (e-mail~: matthieu.moinard@orange-ftgroup.com; isabelle.amonou@orange-ftgroup.com).}
\thanks{P. Duhamel and P. Brault are with the CNRS, Laboratory of Signals and Systems, Supelec, Gif-sur-Yvette, France (e-mail~: patrice.brault@lss.supelec.fr; pierre.duhamel@lss.supelec.fr).}%
}

\maketitle

%
\begin{abstract}
Video compression has been investigated by means of analysis-synthesis, and more particularly by means of inpainting. The first part of our approach has been to develop the inpainting of DCT coefficients in an image. This has shown good results for image compression without overpassing todays compression standards like JPEG. We then looked at integrating the same approach in a video coder, and in particular in the widely used H264/AVC standard coder, but the same approach can be used in the framework of HEVC. The originality of this work consists in cancelling at the coder, then automatically restoring, at the decoder, some well chosen DCT residual coefficients. For this purpose, we have developed a restoration model of transformed coefficients.

By using a total variation based model, we derive conditions for the reconstruction of transformed coefficients that have been suppressed or altered. The main purpose here, in a video coding context, is to improve the rate-distortion performance of existing coders. To this end DCT restoration is used as an \textit{additional prediction step} to the spatial prediction of the transformed coefficients, based on an image regularization process.
The method has been successfully tested with the H.264/AVC video codec standard.
\end{abstract}

\section{Introduction}
%
Image and video compression techniques generally require a transform process. This transform concentrates the energy of the signal into a small number of coefficients. This is what is done in the JPEG still image compression standard and the H264 and HEVC video standards. Trying to improve the compression in such already optimized environments seemed to be a challenge. We first decided to investigate inpainting methods with the idea to automatically regenerate pixels that had been erased. But the spatial prediction of H264 can hardly be improved. We then concentrated our effort on the residual part of the intra prediction that is DCT coded. The residual still contains coefficients that own a sufficiently important energy so that compression could yield to interesting rate gain. We then made the assumption that some of the DCT-transformed coefficients of this residual could be canceled at the coder then automatically retrieved at the decoder with a minimum loss of information and without additional signalling payload.
This assumption led us to develop a first model of cancellation-restoration of DCT coefficients within the framework of a JPEG image coder.\\
%
%
%
%
Many methods have been investigated to perform an efficient post-processing for restoration. In recent works, the optimal reconstruction problem was considered to improve the quality of decoded
images, by authorizing a transform coefficient to oscillate around
its decoded value. The optimal reconstruction develops a model constrained
by both the necessity of a coefficient to belong to its quantization
bin and a regularization process applied directly to the pixels. In
a JPEG system, the work of Alter et al. \cite{Alter2004}, inspired
by S. Zhong \cite{Zhong1996}, significantly improves both the objective and
visual quality of decoded images (similar works apply to wavelets
in \cite{Chan2000}). To get this result, the regularization takes the form
of a Total Variation (TV) minimization problem. The principle of TV minimization \cite{Rudin1992} is commonly used in image processing. Its main advantage is its ability
to preserve edges due to the piecewise smooth regularization
property of the TV semi-norm.

The regularization process based on the TV semi-norm
is formulated as the minimization of a functional, usually solved using
PDE, and has a high computational complexity. Rudin et al. \cite{Rudin1992}
first proposed a gradient projection method to find a solution. Vogel
and Oman \cite{Vogel1996} described a fixed point algorithm, T.F. Chan and A. Chambolle
\cite{Chan1999,Chambolle2004} proposed a new approach based on Newton's
method and Goldstein and Osher \cite{Goldstein2009} introduced a very fast algorithm
based on Bregman iteration. Many applications in the image processing
scope are based on the TV regularization, like noise reduction \cite{Rudin1992,Chambolle1997},
deblurring \cite{Rudin1994,Chan1998}, local inpainting \cite{Chan2002},
zoom-in \cite{Chan2001a}, error concealment \cite{Bourdon2005} and
image compression \cite{Chan2002b}. Although they relate to image
compression, the goal and the methods used are different from the
ones in this paper. T.F. Chan et al. \cite{Chan2002b} introduce a
simple algorithm without transform step: pixels near edges are transmitted,
which requires both their position and value, the others are interpolated
by using TV minimization principle. In view of the amount of information
needed and the property of reconstruction of the TV regularization,
this technique is profitable for specific natural images which have
few edges and large smooth area.

In contrast to these methods, working in a transform domain, like the DCT domain, rather
than in the pixel domain, changes the nature of the inpainting problem,
since one damaged coefficient can affect many pixels. Therefore, geometric
interpolation techniques to restore an image in the pixel domain are
not directly applicable because of their impact on several coefficients.
Direct interpolation in the DCT domain is also
problematic, since DCT coefficients are highly decorrelated.
%
%
For this reason, in this paper we introduce a model based on consecutive switching
between the pixel domain and the DCT domain.
With this formulation, the closest related works seem to be the one
from Alter et al. \cite{Alter2004,Alter2005} based on TV regularization
and the ones from other contributors in error concealment domain \cite{Wang1991,Wang1993,Park1997}
and more generally in \cite{Salama1998}. Indeed, their work intends
to improve the image quality at the decoder side from a degraded (error
concealment) or non-degraded (optimal reconstruction) image. In both
cases, they can be understood as post-processing methods. In contrast,
this paper introduces a new prediction method to be implemented at
both the encoder and the decoder side with the goal to improve the
compression rate. This result is based on a theoretical formulation
of the TV regularization problem in the pixel domain from DCT coefficients.
This work has also been inspired by \cite{Chan2006}, where a new
inpainting method is presented to restore missing wavelet coefficients
of encoded images.

In this paper, we transfer the framework
of \cite{Chan2006} to the B-DCT context and first test it into the framework of a JPEG coder.
The B-DCT is widely used because of its relative ease of implementation and is used
in most present image and video coding standards, as JPEG \cite{Pennebaker1992},
H.264/AVC \cite{Richardson}, SVC \cite{Schwarz2007} and HEVC \cite{ITU-T_JCTVC}. This is
why this paper concentrates on B-DCT.

\begin{figure}[h]
\centering\includegraphics[width=1\linewidth]{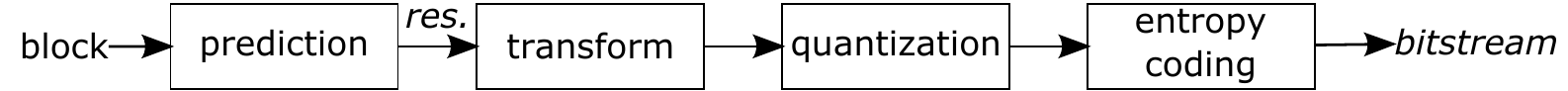}
\caption{Steps to encode a pixel block into a bistream in H264/AVC and HEVC.}
\label{coding_schema}
\end{figure}

%
%
After severals tests on JPEG, we sought to introduce our DCT prediction model as an additional prediction step in a video coding context, namely the video standard H264/AVC. This one implements a block-based codec that includes motion compensation (inter-frame coding) and spatial prediction
methods (intra-frame coding). To be encoded in a bitstream, a block
of pixels follows four steps as illustrated on \green{Fig.}\ref{coding_schema}.
First, prediction is applied in the spatial domain because the goal
is to find pixel similarities between or inside frames. For an original
pixel block $u_{B}$ to code, where $B$ indicates the block position
$u_{B}$ inside the original frame $u$, a predictor block $p$ is
computed using previously encoded frames (inter prediction) or by
an interpolation of the surrounding known pixels (intra prediction). The residue $r$ is expressed as a pixel difference between $p$ and the original block $u_{B}$, as we have $u_{B}=p+r$.
Then a block-based DCT is used to decorrelate the residual signal and to concentrate its energy into
a few number of coefficients. We will also see in section III that block partitioning reveals to be quite interesting to derive a simple regularization term.
%
%
We thus have inserted our DCT coefficients prediction method in the residue of the initial, spatial, prediction stage. In this environment our additional prediction is called \textit{``Visual Coding Residual Prediction'' (VCResPred}). The VcResPred stage thus performs an additional prediction subsequently to the spatial prediction modes (the nine directional modes of H264 here). Clearly, this additional prediction aims at reducing the variance of the global
prediction error, hence at improving the rate/distortion tradeoff. To this end we voluntarily delete, at the encoder side, some specific, predictable, DCT coefficients (those that can be efficiently predicted, i.e. restored).
%
%
Then the difference between the predicted DCT values and the actual ones is computed, resulting in a prediction
error which is then quantized and coded. At the decoder side, the
same prediction of the DCT residual coefficients is done and the block
is correctly reconstructed by adding coefficient predictions and residual
errors. This DCT-prediction mechanism turns out to be very similar to the
ones actually found in H264/AVC for the fact
that we add a prediction on top of another, temporal and/or spatial,
prediction.
%
%
%
%

%
%
The outline of this paper is as follows~: our compression problematic and state-of-art related to inpainting, total variation based regularization and DCT coefficients restoration are introduced in section I.
%
%
In section II we present our Total Variation based regularization model for the prediction of DCT coefficients.
In section III, we introduce a cancellation/restoration model, based on the regularization model of section II, for the prediction of DCT coefficients in the context of still image coding, and show the results on a JPEG coder. In section IV we derive this model for the integration in a video coding context, and more specifically H264/AVC, with emphasizing the necessary modifications for an effective integration, and show the results in this context. We conclude, in section V, with a summary of this work, of its results and of prospective improvements.

\section{Regularization model for the restoration of DCT coefficients in a
prediction based image and video coder}\label{Section Regularization Model}

In this section, the goal is to link the Total Variation minimization problem in the
pixel domain with coefficients-to-predict in the DCT domain. From there,
a model of DCT coefficients restoration by TV minimization in pixel
domain could ensue. Rather than explaining our model first in an image context and then in a video context, we make the choice here to only describe the regularization model in the video context. The image model can be derived in a straightforward manner, from the following development, by assuming that the initial spatial prediction, of the video coder, $p_{\zvec}$ is NULL and that the DCT residual (coefficients) $r_{\kvec}$ of the video context corresponds to the principal (only) coefficients of an image context. Here $\zvec$ stands for a coordinates vector in the pixel domain and $\kvec$ for the coordinates in the DCT domain.

In a video coding context like H264 or HEVC, a block $u_{B}$ is defined from the sum of a predicted block $p_{\zvec}$ of a set of pixels (spatial prediction), and a residue $r_{\kvec}$ of DCT coefficients. With this notation, each block
$u_{B}$ is treated independently. So we have the relation:

\begin{equation}\label{fundamental_video_prediction_relation}
u_{\vec{z}B}=p_{\vec{z}}+r_{\vec{k}}\phi_{\vec{z},\vec{k}}
\end{equation}

So, the TV for a given block $u_{B}$ is now given by :

\begin{equation}
TV\left(u_{\vec{z}B}(p_{\vec{z}},r_{\vec{k}})\right)=\int_{\Omega}|\nabla_{\vec{z}}u_{\vec{z}B}(p_{\vec{z}},r_{\vec{k}})|d\vec{z},\label{eq:tv(u)}
\end{equation}
where $\Omega$ is defined on $\mathbb{R}^{2}$ . Then, to minimize
the TV, we compute its partial differential equation:

\begin{eqnarray}\begin{split}
\frac{\partial TV(u_{\vec{z}B}(p_{\vec{z}},r_{\vec{k}}))}{\partial r_{\vec{k}}}
=\int_{\Omega}\frac{\partial|\nabla_{\vec{z}}u_{\vec{z}B}(p_{\vec{z}},r_{\vec{k}})|}{\partial r_{\vec{k}}}\mathrm{d}\vec{z}\nonumber \\
= \int_{\Omega}\frac{\nabla_{\vec{z}}u_{\vec{z}B}(p_{\vec{z}},r_{\vec{k}})}{|\nabla_{\vec{z}}u_{\vec{z}B}(p_{\vec{z}},r_{\vec{k}})|}\cdot\frac{\partial\nabla_{\vec{z}}u_{\vec{z}B}(p_{\vec{z}},r_{\vec{k}})}{\partial r_{\vec{k}}}\mathrm{d}\vec{z}\nonumber \\
=
\int_{\Omega}\frac{\nabla_{\vec{z}}u_{\vec{z}B}(p_{\vec{z}},r_{\vec{k}})}{|\nabla_{\vec{z}}u_{\vec{z}B}(p_{\vec{z}},r_{\vec{k}})|}\cdot\nabla_{\vec{z}}\frac{\partial u_{\vec{z}B}(p_{\vec{z}},r_{\vec{k}})}{\partial r_{\vec{k}}}\mathrm{d}\vec{z}\label{eq:TV_derivee _partielle}
\end{split}\end{eqnarray}

\vspace{.2cm}
From the definition of the inverse DCT, we have~:
\begin{equation}
\frac{\partial u_{\vec{z}B}(p_{\vec{z}},r_{\vec{k}})}{\partial r_{\vec{k}}}=\phi_{\vec{z},\vec{k}}\label{eq:simplify_to_phi}
\end{equation}
where $\phi_{\vec{z},\vec{k}}$ is the DCT kernel. So, the new formulation
gives :
\begin{equation}
\frac{\partial TV(u_{\vec{z}B}(p_{\vec{z}},r_{\vec{k}}))}{\partial r_{\vec{k}}}=\int_{\Omega}\frac{\nabla_{\vec{z}}u_{\vec{z}B}(p_{\vec{z}},r_{\vec{k}})}{|\nabla_{\vec{z}}u_{\vec{z}B}(p_{\vec{z}},r_{\vec{k}})|}\cdot\nabla_{\vec{z}}\phi_{\vec{z},\vec{k}}\mathrm{d}\vec{z}.
\end{equation}

An integration-by-parts yields :
\begin{equation}
\begin{split}
\int_{\Omega}\frac{\nabla_{\vec{z}}u_{\vec{z}B}(p_{\vec{z}},r_{\vec{k}})}{|\nabla_{\vec{z}}u_{\vec{z}B}(p_{\vec{z}},r_{\vec{k}})|}\cdot\nabla_{\vec{z}}\phi_{\vec{z},\vec{k}}\mathrm{d}\vec{z}=\\
\left[\nabla_{\vec{z}}\cdot\left(\frac{\nabla_{\vec{z}}u_{\vec{z}B}(p_{\vec{z}},r_{\vec{k}})}{|\nabla_{\vec{z}}u_{\vec{z}B}(p_{\vec{z}},r_{\vec{k}})|}\phi_{\vec{z},\vec{k}}\right)\right]_{0}^{\Omega}\\
-\int_{\Omega}\left(\nabla_{\vec{z}}\cdot\frac{\nabla_{\vec{z}}u_{\vec{z}B}(p_{\vec{z}},r_{\vec{k}})}{|\nabla_{\vec{z}}u_{\vec{z}B}(p_{\vec{z}},r_{\vec{k}})|}\right)\phi_{\vec{z},\vec{k}}\mathrm{d}\vec{z},
\end{split}
\label{eq:int_par_partie}
\end{equation}

Due to the block partitioning (for the DCT) of the image, the DCT kernel $\phi_{\vec{z},\vec{k}}$
is zero outside the block $u_{B}$. Then :
\begin{equation}
\left[\nabla_{\vec{z}}\cdot\left(\frac{\nabla_{\vec{z}}u_{\vec{z}B}(p_{\vec{z}},r_{\vec{k}})}{|\nabla_{\vec{z}}u_{\vec{z}B}(p_{\vec{z}},r_{\vec{k}})|}\phi_{\vec{z},\vec{k}}\right)\right]_{0}^{\Omega}=0.
\end{equation}

Finally, the partial differential equation of the total variation
gives~:
\begin{equation}
\begin{split}
& \frac{\partial TV(u_{\vec{z}B}(p_{\vec{z}},r_{\vec{k}}))}{\partial r_{\vec{k}}}=\\
&\qquad -\int_{\Omega}\left(\nabla_{\vec{z}}\cdot\frac{\nabla_{\vec{z}}u_{\vec{z}B}(p_{\vec{z}},r_{\vec{k}})}{|\nabla_{\vec{z}}u_{\vec{z}B}(p_{\vec{z}},r_{\vec{k}})|}\right)\phi_{\vec{z},\vec{k}}\mathrm{d}\vec{z}.\label{eq:partial_differential_TV}
\end{split}
\end{equation}

The term $\nabla\cdot\left[\nabla u/|\nabla u|\right]$ is an expression
of curvature \cite{Chan2001} \textit{}expressed in the pixel domain.
This formula connects geometric information in the spatial domain
with the DCT kernel $\phi_{\vec{z},\vec{k}}$, hence characterizes
the TV regularization constraint in the DCT domain.

Returning to the original problem, the model consists in the TV minimization
for specified DCT coefficients $\vec{k}\in\mathcal{I}_{DCT}$ :
\begin{equation}
\min_{r_{\kvec},\; \kvec \in\mathcal{I}_{DCT}} TV(u_{\zvec B}(p_{\vec{z}},r_{\kvec}))
\end{equation}

For $TV(u_{\vec{z}B}(p_{\vec{z}},r_{\vec{k}}))=0$, the associated
Euler-Lagrange equations gives :
\begin{equation}
\frac{\partial TV(u_{\vec{z}B}(p_{\vec{z}},r_{\vec{k}}))}{\partial r_{\vec{k}}}=0
\end{equation}

Finally, with Eq. \ref{eq:partial_differential_TV} , we can express
the TV minimization problem in the DCT transformed domain. Then, we
are able to regularize an image by varying the corresponding DCT coefficients
of the residue.

\section{Implementation 1 - Cancellation/Restoration of DCT coefficients in an image\label{sub:Inpainting-algorithm}}

\subsection{Description of the method}

The algorithm we introduce aims at restoring some DCT coefficients
$r_{\vec{k}}$ for all $\vec{k}\in\mathcal{I}_{DCT}$, where $\mathcal{I}_{DCT}$ is the subset of the position
of missing DCT coefficients.

We first assume that the position of the coefficients to restore,
$\mathcal{I}_{DCT}$, is known. In order to experiment the inpainting
process, we also assume to be in the special case where the prediction
$p_{\vec{z}}$ is null. In this case, the residue $r_{\vec{k}}$ is composed
of the block pixels directly transformed in the DCT domain. Actually,
removing the prediction step, we are coming back to a basic image
coder as JPEG without spatial prediction.

The first step is to compute the image from the DCT coefficients,
so :
\begin{equation}
u_{\vec{z}B}=p_{\vec{z}}+ DCT^{-1}(r_{\vec{k}}),\ \forall u_{B}\in u
\end{equation}

where $DCT^{-1}$ represents the inverse DCT. As $p_{\vec{z}}$ is null,
we simply formulate $u_{\vec{z}B}=DCT^{-1}(r_{\vec{k}})$.

Then, the curvature, which is the guideline of the coefficient inpainting
update, is projected in the DCT domain :

\begin{equation}
c_{\vec{k}}=DCT\left(curv(u_{\vec{z}B})\right),
\end{equation}

where $c_{\vec{k}}$ is the convergence term of the algorithm and
$curv$ the curvature function. The method described in \cite{Chan2002b}
is used to compute the curvature of a discrete function. Then, $c_{\vec{k}}$
allows to update corresponding DCT coefficients to restore $\vec{k}\in\mathcal{I}_{DCT}$
of $r_{\vec{k}}$ so that the total variation of $u_{B}$ is reduced. By an iterative process, the algorithm tends to a local minimum. The step $\gamma_{i}$ is function of the number of iterations $i$
of the algorithm.
%
%
The stop conditions of the algorithm are defined by the maximum number
of iterations $L$, and by the parameter $\delta$ which indicates
if a stationary state is reached. This algorithm is used to restore missing DCT coefficients $\vec{k}\in\mathcal{I}_{DCT}$.

\subsection{Experiments and results}

$\bullet$ First, we simulate a loss of information by randomly canceling a
percentage of DCT coefficients in a B-DCT coded image. Our algorithm recovers these
coefficients in the DCT domain, by the regularization process above. This result was presented in \cite{Moinard011}
%
%

$\bullet$ Second, in order to test our method on optimal image decoding/reconstruction, we conduct the subsequent experiment where no coefficient is removed but the image is quantized. The goal is to suppress the decoding artifacts due to the quantization. In this experiment we have obtained an average gain of $0.7$dB in PSNR.

%
$\bullet$ Third, in order to develop a new coding method, we sought to characterize the
quality improvement illustrated in Fig. 1 of \cite{Moinard011}
with a notion of information amount, depending on a specific position
$\vec{k}=(i,j)$ of a DCT coefficient in a $8\times8$ block, with
$0 \leq  i,j < 8$. By doing so we wish to highlight
the position $(i,j)$ of the transformed coefficients that can be
well restored and that significantly reduce the entropy when suppressed.

$\bullet$ Following this idea, we first introduce our prediction model, as a new prediction method, into a JPEG coding scheme. Each image block $u_B$ undergoes a DCT in order to get the corresponding DCT block $\alpha_B$. The DCT $\alpha_{B\kvec}$ coefficients are split into two complementary sets $\mathcal{I}_O$ and $\mathcal{I}_{DCT}$, with the coefficients of the first set being used as support for the prediction of the second set coefficients.
To optimally define the $\mathcal{I}_{DCT}$ set, it must correspond to the coefficients that~:\\
- can be correctly predicted. \\
- own a non negligeable energy.\\
These two criteria are conflicting and we adopt a compromise principle of rate/distorsion. Two ways are possible~: use a static, predefined, $\mathcal{I}_{DCT}$ set, or compute a dynamic set in which we have to transmit the predicted coefficients configuration to the decoder. With regard to our first experiments, we took the first choice  which is to have NO overhead information to transmit.

\begin{figure}[h]
\centering\includegraphics[width=1\linewidth]{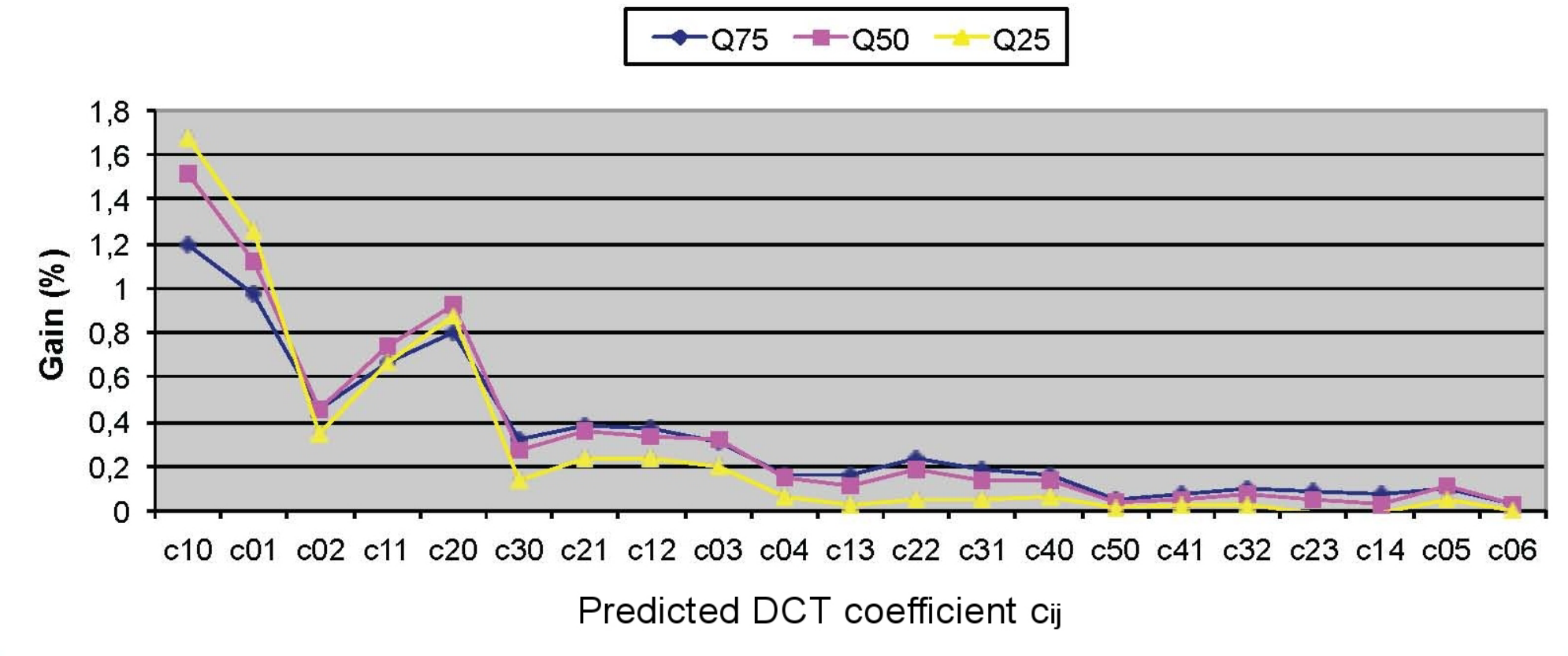}
\caption{Mean rate gain (ref~: JPEG) vs only one predicted $c_{ij}$ coefficient and a coding quality $Q\in[25,50,75]$.}
\label{Fig:JPEG-result}
\end{figure}

%
%
$\bullet$ At this point, a first experiment was to delete the
coefficients from a \textit{specific position} $(i,j)$ (see \green{Fig.} \ref{Fig:JPEG-result}) of all DCT blocks of
the image. The other coefficients undergo a scalar quantization by
using the standard quantization matrix from JPEG. The best result was obtained for $\mathcal{I}_{DCT}=[c_{10}]$ and also for low-rates, with a gain of $1.68\%$.

\begin{figure}[h]
\centering\includegraphics[width=1\linewidth]{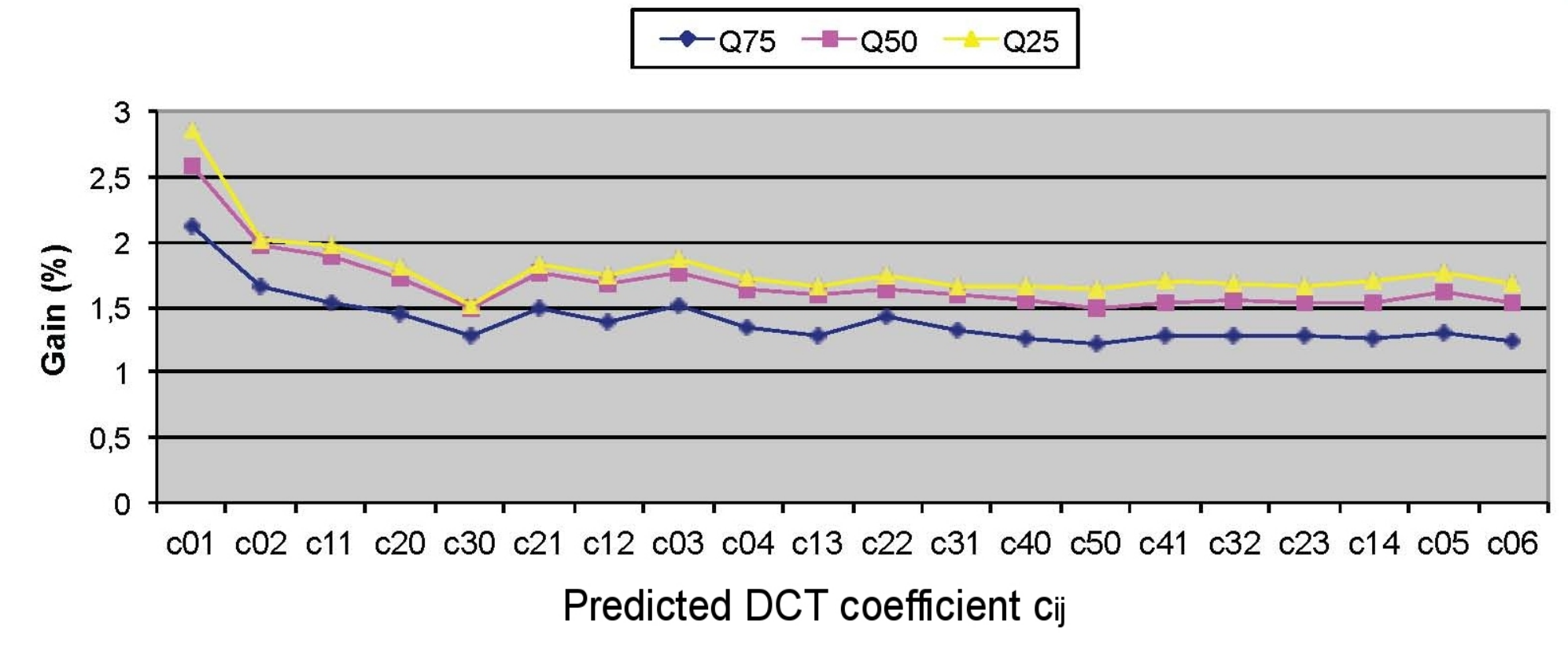}
\caption{Mean rate gain (ref~: JPEG) vs two predicted coefficients, one fixed $(c_{10})$ and a second variable $c_{ij}$, i.e. with $\mathcal{I}_{DCT}=[c_{10},c_{ij}]$, and a coding quality $Q\in[25,50,75]$.
}
\label{Fig:JPEG-result-c10}
\end{figure}

$\bullet$ A second experiment was to suppress several coefficients. We then fixed $\mathcal{I}_{DCT}=[c_{10},c_{ij}]$ (see \green{Fig.} \ref{Fig:JPEG-result-c10}). As might be expected, the deleted coefficients
which induce a maximum entropy reduction are those which mostly reduce
the image quality (coefficients $(1,0)$ and $(0,1)$). This experiment
tells us that in order to reduce the amount of information for an efficient image coding, it is necessary to delete the DCT coefficients corresponding to the low frequencies, because others have no effect on the entropy.
In fact, these are already set to null by the quantization process. We thus observed the maximum gain of $2.51\%$ for $\mathcal{I}_{DCT}=[c_{10},c_{01}]$. By repeating this experiment with more coefficients, we noticed that the gain was no longer improved.

\begin{figure}[h]
\centering \includegraphics[width=1\linewidth]{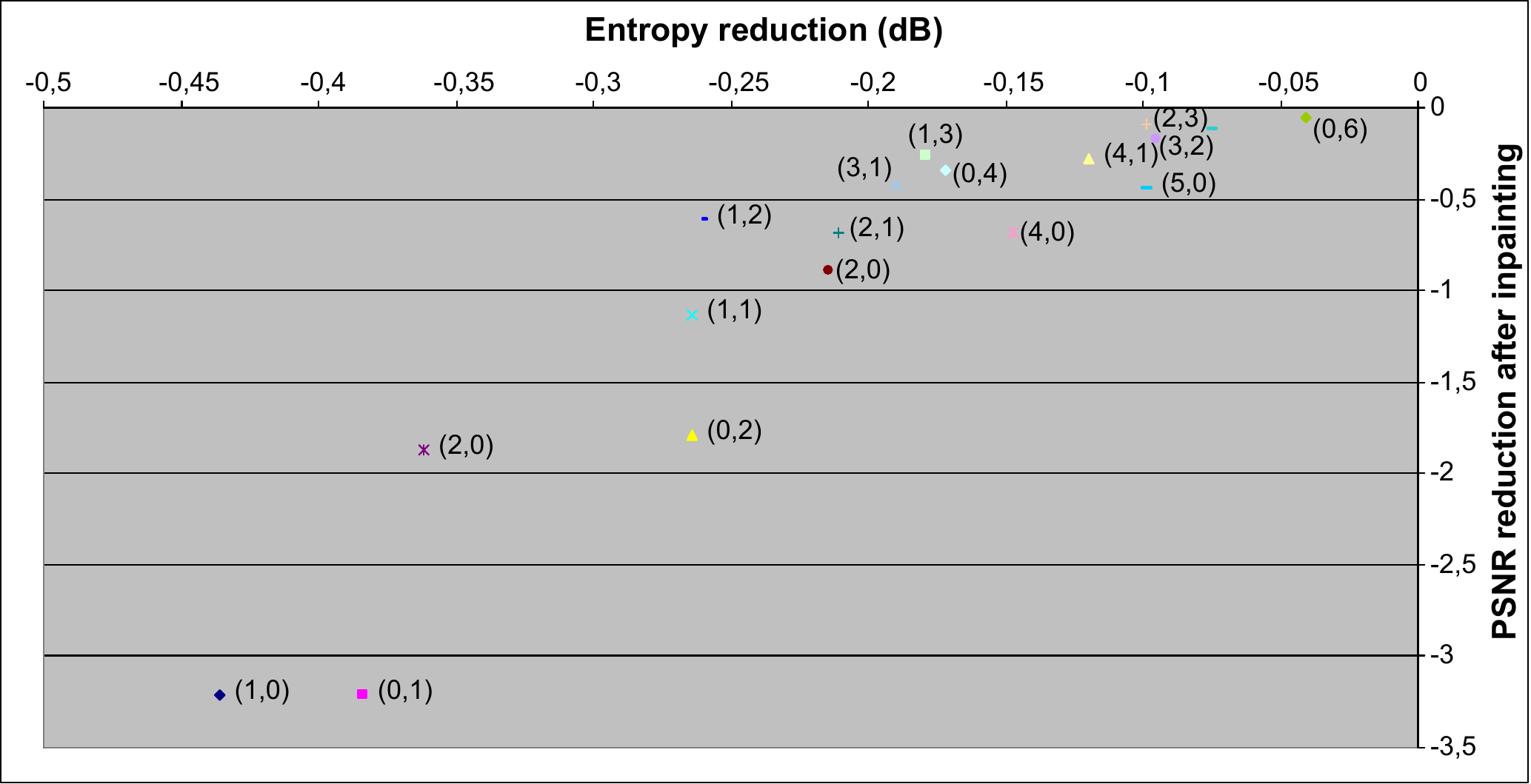}
\caption{{Entropy reduction and PSNR reduction for a specific position $\vec{k}=(i,j)$
in a DCT block of deleted/restored coefficients with the image boat. This figure demonstrates that it is the first, low frequency coefficients of the DCT (after the DC coeff.) that present the highest entropy reduction, thus the highest interest for compression, but also the highest reduction of PSNR after restoration.}}
\label{Fig:Entropy_vs_PSNR}
\end{figure}

%
%
\green{Fig.} \ref{Fig:Entropy_vs_PSNR} illustates the entropy reduction according
to the PSNR reduction after the inpainting process for some different
positions $(i,j)$ of DCT coefficients. For the other positions, their
impact on entropy and PSNR reduction are negligible, so they are not
illustrated on this graphic.
%
%
%
%
%
%
\begin{figure}[h]
\centering%
\begin{tabular}{cc}
\subcaptionbox{Sub-image of the original image ``boats''}
{\includegraphics[width=0.45\linewidth]{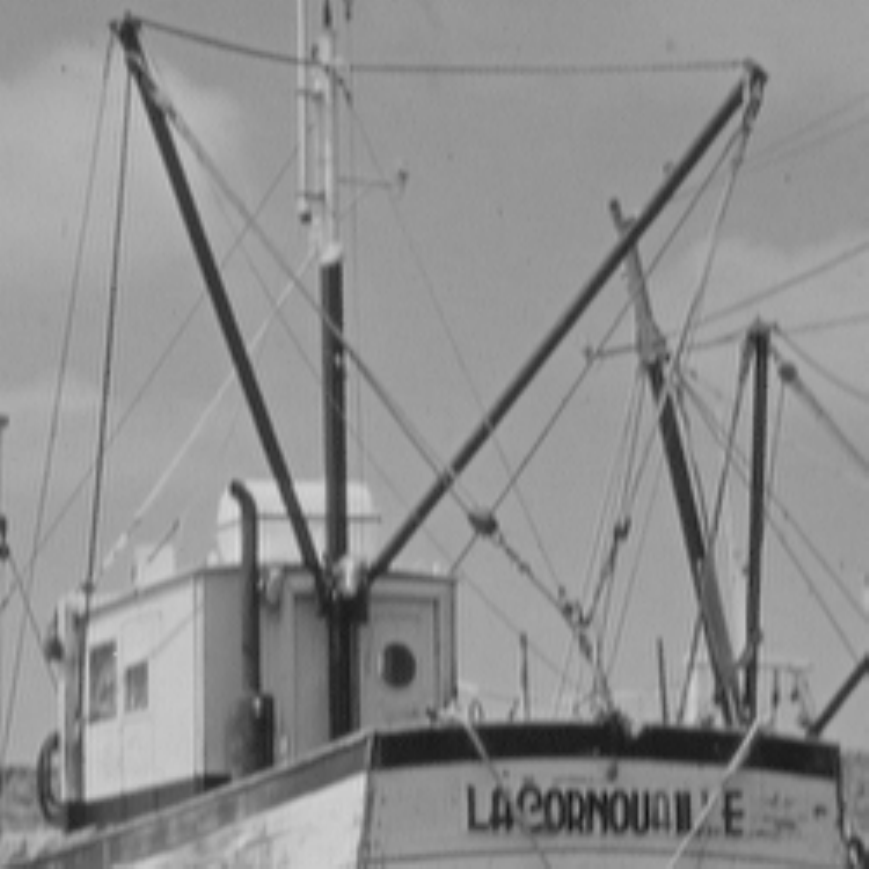}
\label{Fig:JPEGEX_or}}
& \subcaptionbox{Step $1$: decoding of $\mathcal{I}_{O}$ ($PSNR=24,5\; dB$)}
{\includegraphics[width=0.45\linewidth]{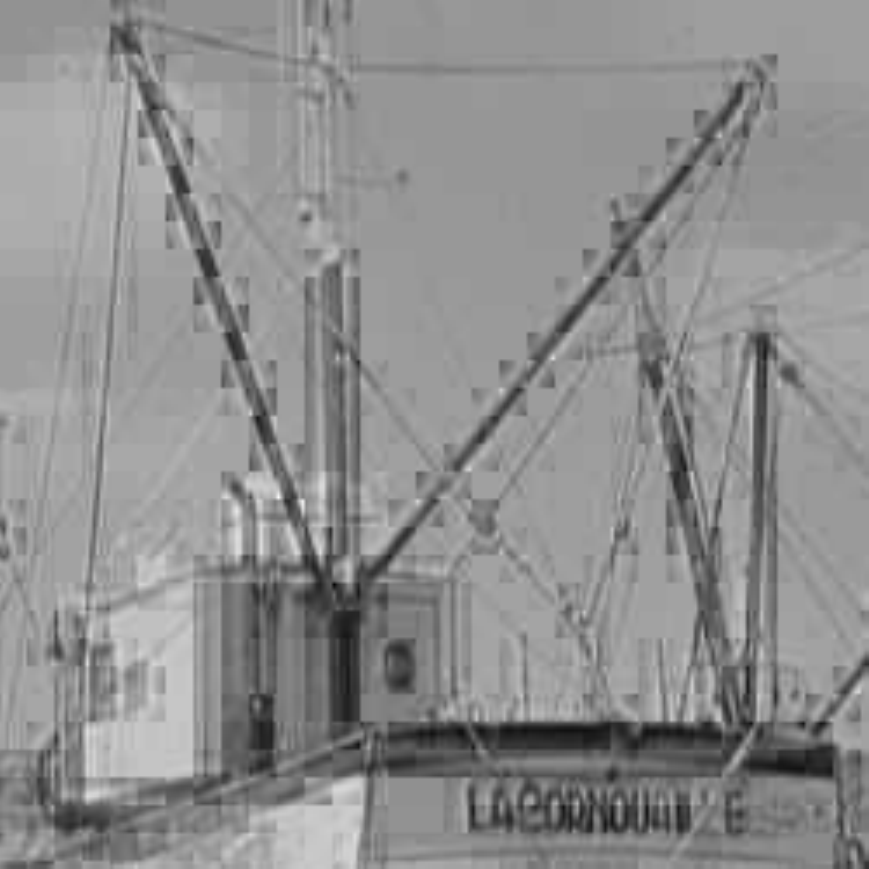}
\label{Fig:JPEGEX_first}}\tabularnewline
\subcaptionbox{Step $2$: prediction of $I_{DCT}$ ($PSNR=28,15\; dB$)}
{\includegraphics[width=0.45\linewidth]{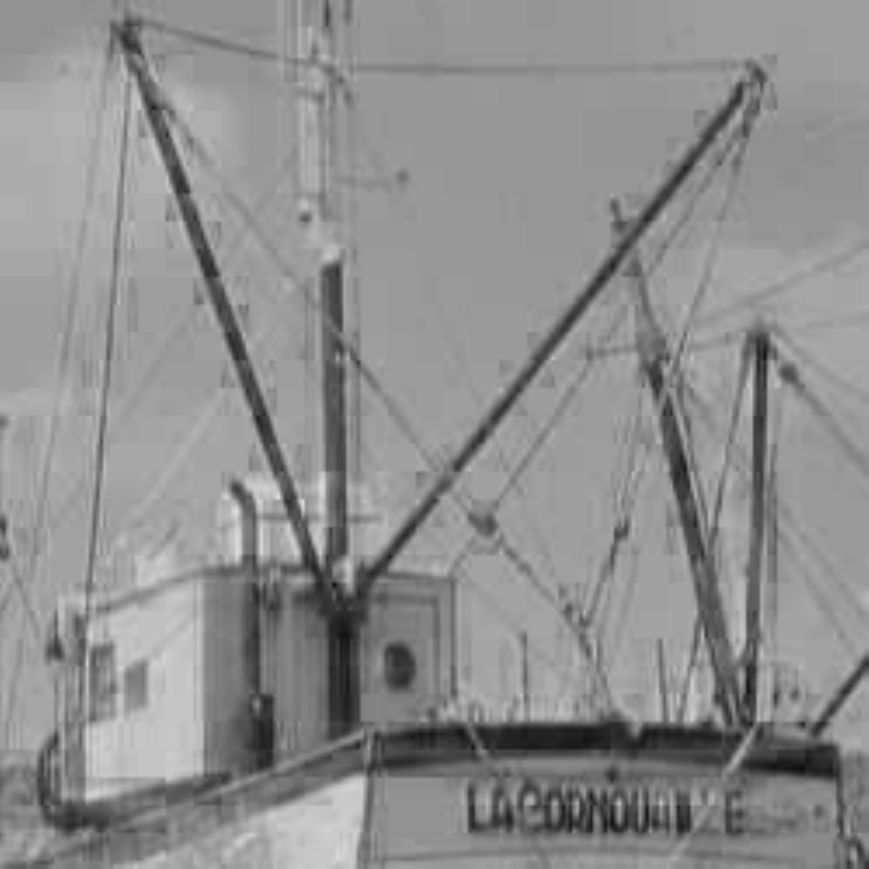}
\label{Fig:JPEGEX_inpaint}}
& \subcaptionbox{Step $3$: adding the prediction error for $I_{DCT}$ ($PSNR=34,68\; dB$)}
{\includegraphics[width=0.45\linewidth]{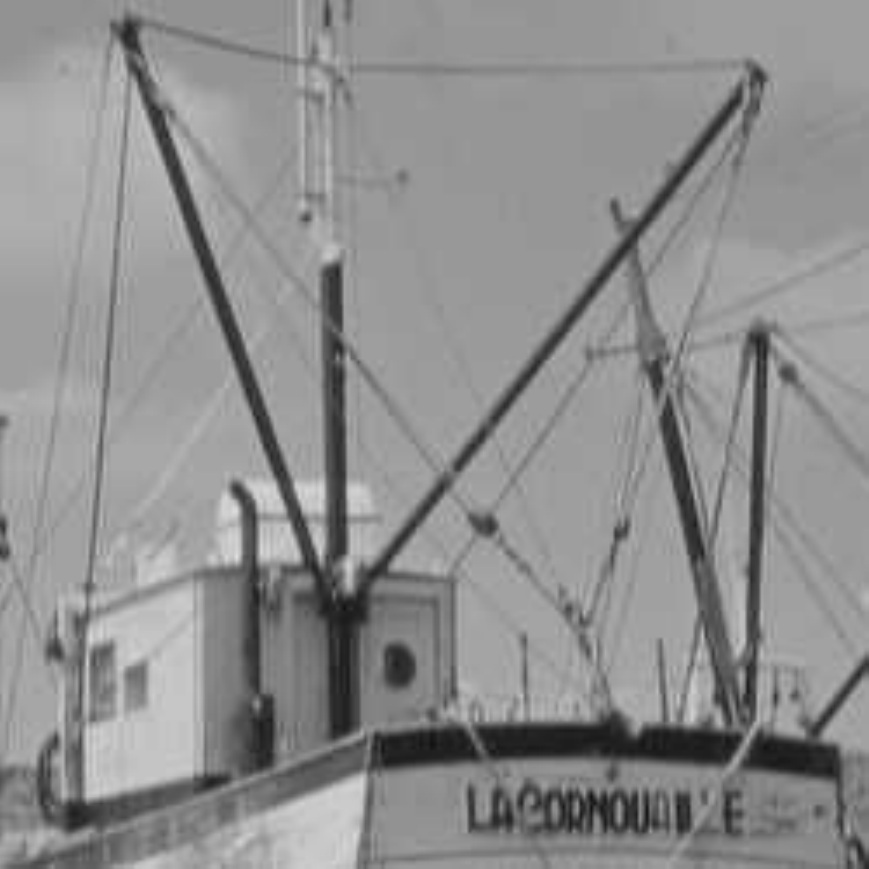}
\label{Fig:JPEGEX_final}}\tabularnewline
\end{tabular}
\caption{Illustration of the decoding steps of our compression algorithm based on the prediction of DCT coefficients and inserted in a JPEG codec (B-DCT + quantization). The predicted DCT coefficients are $(1,0)$ and $(0,1)$.}
\label{Fig:JPEGEX}
\end{figure}

\begin{table}[h]
\centering%
\begin{tabular}{|c||c|c||c|c|}
\cline{2-5}
\multicolumn{1}{c|}{} & \multicolumn{2}{c||}{$Q75$} & \multicolumn{2}{c|}{$Q25$}\tabularnewline
\cline{2-5}
\multicolumn{1}{c|}{image} & rate (bpp) & $\Delta\;(\%)$  & rate (bpp) & $\Delta\;(\%)$\tabularnewline
\hline
\hline
\textit{barbara} & $1,32$ & $2,24$  & $0,56$ & $2,88$\tabularnewline
\hline
\textit{boats} & $0,95$ & $2,19$  & $0,39$ & $3,44$\tabularnewline
\hline
\textit{bridge} & $1,87$ & $1,47$  & $0,74$ & $2,27$\tabularnewline
\hline
\textit{couple} & $1,29$ & $2,02$  & $0,52$ & $2,62$\tabularnewline
\hline
\textit{crowd} & $1,22$ & $3,27$  & $0,54$ & $4,14$\tabularnewline
\hline
\textit{dollar} & $2,14$ & $0,59$  & $0,87$ & $0,73$\tabularnewline
\hline
\textit{girlface} & $0,81$ & $3,82$  & $0,3$ & $5,24$\tabularnewline
\hline
\textit{kiel} & $1,48$ & $1,01$  & $0,59$ & $1,44$\tabularnewline
\hline
\end{tabular}

\caption{Our inpainting-based, two-DCT coefficients ($[1,0]$ and $[0,1]$), prediction method in JPEG~: results per image and per rate (ref~: JPEG). The rate obtained is expressed in bits/pixel. The gain w.r.t. JPEG is expressed in $\%$
}
\label{Tab:JPEG-resultats}
\end{table}
\section{Implementation 2 - prediction in a DCT based video coding}

By relying on our regularization model and our former experiments on JPEG, a new prediction method is integrated into an MPEG-4 AVC/H264 video encoder. We now show that a substantially different strategy has to be adopted for the predicted coefficients choice.

\subsection{New prediction method for the DCT coefficients}

The proposed coder works in two stages. First, the DCT coefficients
of the residue are divided to select those which are deleted. Then,
the prediction process is applied to restore the coefficients previously
deleted. The prediction error is computed and sent to the entropy
coder. The layout of the modified video encoder is introduced on \green{Fig.} \ref{Fig:new_encoder_schema}. $\mathcal{I}_{DCT}$ is the set of the DCT coefficient positions in a block, which are predicted using
our proposed method. Notice that if $\mathcal{I}_{DCT}$ is empty,
the encoding process is exactly similar to a classical
encoder (H264, HEVC) \cite{Richardson,ITU-T_JCTVC,Sullivan2012}: prediction, transform, quantization and entropy coding.


The prediction process of DCT residual coefficients, that we name \textit{VCResPred}, is applied on each block individually. This process needs to be the same at the encoder and the decoder sides. The DCT
residual $r_{\kvec}$, i.e. the DCT-coded (spatial intra-inter) prediction error, is split between the coefficients
$\mathcal{I}_{DCT}$ we want to predict and the unmodified ones, called $\mathcal{I}_{O}$
(original coefficients). An advantage of this method is that some
of the valid information inside a block (coefficients not in $\mathcal{I}_{DCT}$)
is used to predict the remainder ($r_{\vec{k}}$ for all $\vec{k}\in\mathcal{I}_{DCT}$),
so it allows an intra-block prediction. From this point of view, the
method is different from the block prediction in pixel domain
where the entire block is predicted whenever. Moreover, we can accurately tune the set $\mathcal{I}_{DCT}$ in order to find the optimal rate/distorsion compromise. Following experiments will show how to
adapt $\mathcal{I}_{DCT}$ according to the intra prediction mode.

For original coefficients $\mathcal{I}_{O}$ which are used to predict
the $\mathcal{I}_{DCT}$, they are quantized then de-quantized at the
encoder side since they need to be identical at the encoder and the
decoder side. We also require the coded/decoded causal neighbor blocks
to compute the curvature in the spatial domain for the current block
$u_{\vec{z}B}$. For all $\vec{k}\in\mathcal{I}_{DCT}$, $r_{\vec{k}}$
are first set to zero and then predicted. The residue of the DCT prediction
is obtained by taking the difference between the original DCT coefficient
and the predicted coefficient. Finally, the \textit{VCResPred} prediction
error is quantized and coded.

The main difficulty is to \textbf{find the optimal set} $\mathcal{I}_{DCT}$
of the DCT coefficients to predict (and the complementary set $\mathcal{I}_{O}$)
. Ideally, $\mathcal{I}_{DCT}$ must correspond to the DCT coefficients
:
\begin{itemize}
\item that can be correctly predicted using the inpainting algorithm previously introduced.
\item that have significant energy.
\end{itemize}
This second point is crucial if we want the method to reduce the entropy
of coded information and to improve the compression rate. We also must remark that, unlike the coding scheme formerly proposed for still images, we cannot adopt the two static coefficients prediction framework. This is mainly due to the fact that the spatial prediction reduces the energy of the residue w.r.t. the original information. And because we have previously noticed that most of the energy is embedded in the two low-frequency DCT coefficients, we have to adopt a different strategy for video residual coefficients prediction than in the case of still image. This is what we are going to explore now by declining two important specificities encountered by our prediction approach in the case of video compression, namely i) the spatial prediction mode inference on the choice of DCT predicted coefficients and ii) the choice of an iterative prediction for the improvement of the method and for the integrity of the coder/decoder synchronization.

\begin{figure}[h]
%
\centering\includegraphics[width=1\linewidth]{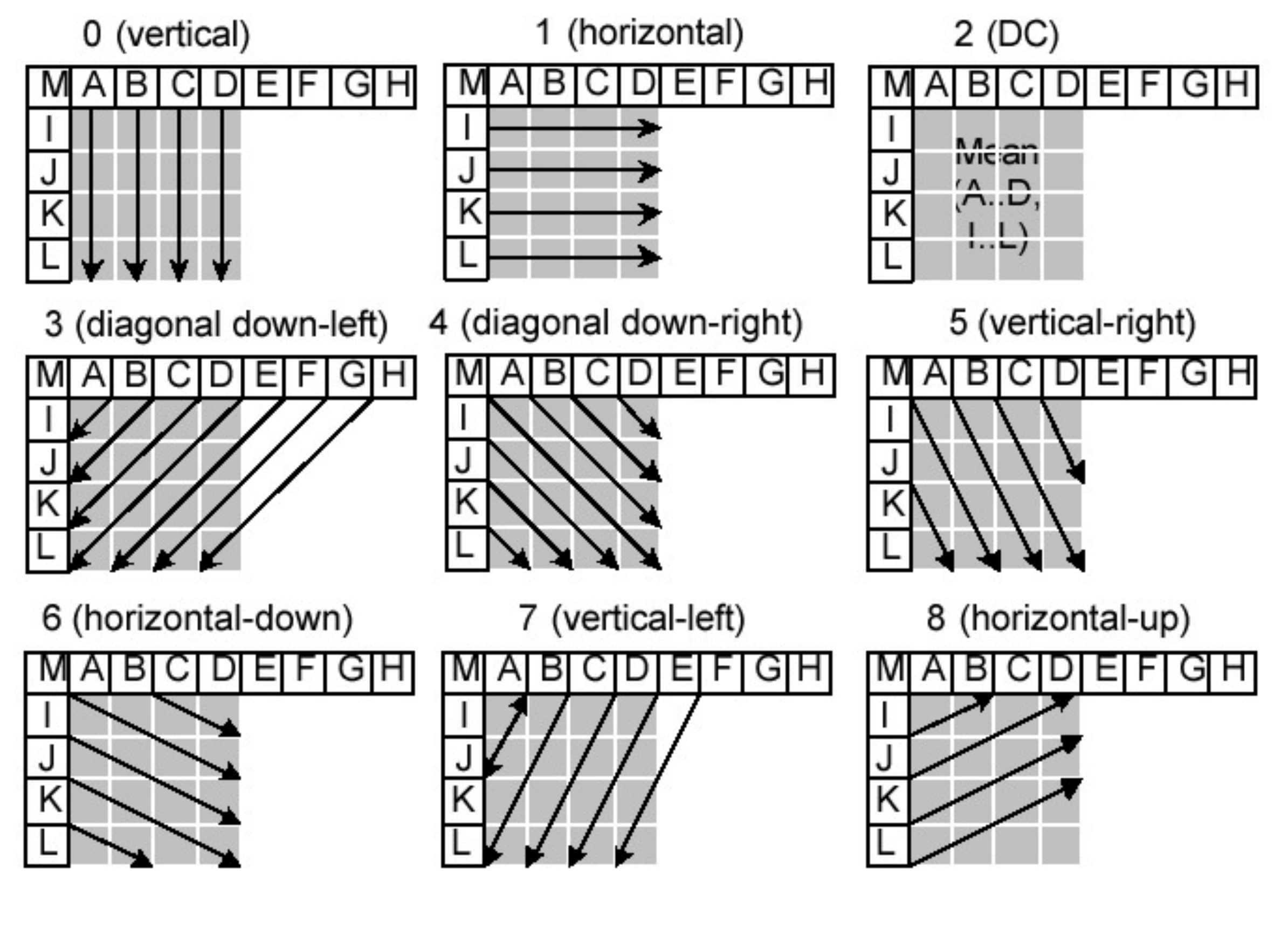}
\caption{The nine intra prediction modes of the H264 standard \cite{Richardson}, increased in HEVC to the 33 angular modes, up to the $1/8$th pel accuracy, plus the planar prediction.
Pixels from $A$ to $M$ are known pixels.}
\label{Fig:H264_intra_pred_mode}
\end{figure}

In the rest of this paper, we will focus on the intra-frame prediction
of H264/AVC, but the method could be transposed to the inter-prediction
process, as well as to the HEVC video codec.\\
There are nine intra prediction modes in H264 for $4\times4$
and $8\times8$ blocks: eight modes use a pixel interpolation according
to a direction, the last is a block prediction using the mean of the
neighbor pixels (DC mode) (the \green{Fig.} \ref{Fig:H264_intra_pred_mode}
illustrates the nine prediction modes for $4\times4$ blocks).\\

\begin{figure}[h]
\centering\includegraphics[width=0.8\linewidth]{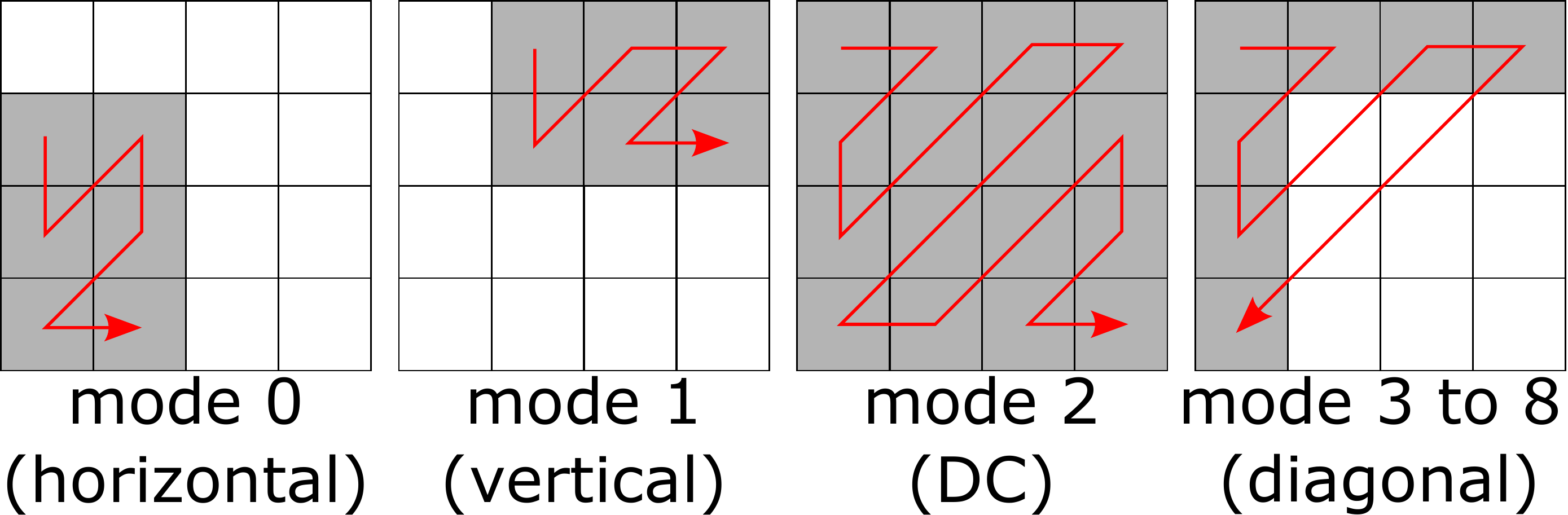}
\caption{Position (in gray) of the predicted coefficients of the $\mathcal{I}_{DCT}$ set for each
$4\times4$ block, according to the intra prediction mode. In red is the prediction order of the coefficients.}
\label{Fig:inpainted_coefficients_4x4}
\end{figure}

\begin{figure}[h]
\centering\includegraphics[width=1\linewidth]{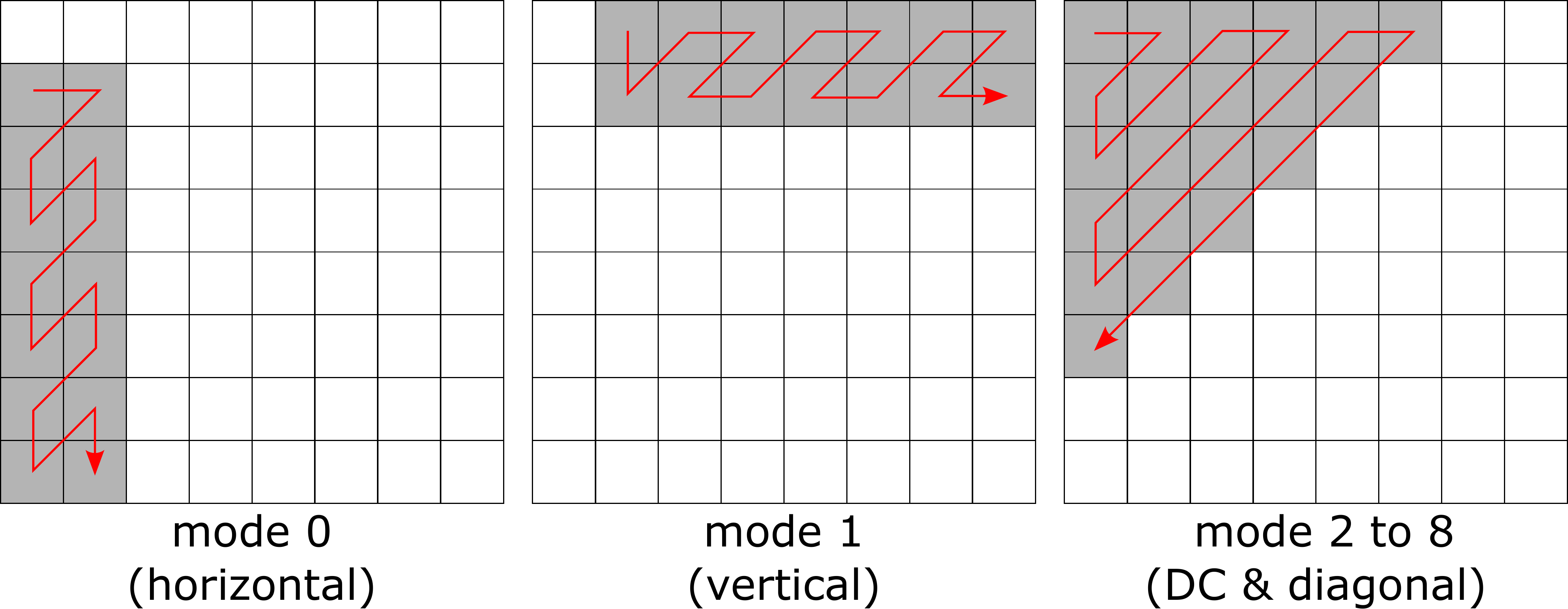}

\caption{Position (in gray) of the $\mathcal{I}_{DCT}$ set for each $8\times8$ block, according to the intra prediction mode. In red is the prediction order of the coefficients.}
\label{Fig:inpainted_coefficients_8x8}
\end{figure}

$\bullet$ A first important point specific to the case of video compression, is to notice \cite{Wu2007} that there is a correlation between the intra prediction modes and the distribution of the DCT coefficients in a residual block. If we notice that an horizontal intra prediction mode reflects a vertical oscillation, we easily deduce that the residual energy is concentrated in the vertical DCT coefficients. It is therefore natural to choose to predict these coefficients. In H264/AVC, the used scan order is the zigzag order \cite{Richardson},
that browses the DCT coefficients from lower to higher frequencies
because of the well-known energy distribution in a DCT block. But with our new residual prediction method it is more efficient to adopt a multiple directions scanning scheme. This way, we locate consecutive zeros at the rear part of the block scan, so that the entropy coder is optimized because of the knowledge of the distribution
of the coefficients. By declining the concept, the set $\mathcal{I}_{DCT}$
is predefined at both encoder and decoder sides according to the intra
prediction mode. For the DC mode, all DCT coefficients of the block
are predicted as illustrated on \green{Fig.} \ref{Fig:inpainted_coefficients_4x4}
and \green{Fig.} \ref{Fig:inpainted_coefficients_8x8} for $4\times4$ and
$8\times8$ respectively. The set $\mathcal{I}_{DCT}$ has been constructed
experimentally for each intra prediction mode.

At the decoder side and with the decoded intra prediction mode, we
directly have the predefined position of block coefficients to predict
$\mathcal{I}_{DCT}$. Then, the DCT prediction is applied on the current
block exactly as at the encoder side. Finally, additioning the DCT
prediction errors with the predicted DCT coefficients, we are able
to reconstruct the DCT block.\\

$\bullet$ The second important point in the case of video compression is now explained. The inpainting algorithm introduced in \ref{sub:Inpainting-algorithm} and now declined in a video coding context was able to predict all the
coefficients from $\mathcal{I}_{DCT}$ of the block simultaneously.
However, it seemed more appropriate to process the coefficients $r_{\kvec}, \; \forall \kvec \in
\mathcal{I}_{DCT}$ iteratively and hierarchically. Indeed, when
a coefficient is predicted and the residue is coded and decoded, it
can be used to predict the other coefficients of $\mathcal{I}_{DCT}$
without breaking the synchronization between the encoder and decoder.
So, the first predicted coefficient of $\mathcal{I}_{DCT}$ is the
one corresponding to the lower frequency of the block. Then, once
the prediction is made, the prediction error is computed and coded,
hence this coefficient can be used to predict the other coefficients
from $\mathcal{I}_{DCT}$ of the block, until the higher frequency
coefficient is reached. The order of the predicted DCT coefficients
is defined by one of the four or three directions scan method.

\subsection{Implementation and results}

The DCT prediction method has been implemented in the JSVM 9.7 reference
software \cite{Schwarz2005d} (without the scalable part), with Context-based
Adaptive Binary Arithmetic Coding (CABAC) entropy coding and $8\times8$
transform turned on. In these experimentations, the frames are only
coded in intra mode. The DCT prediction method is used for $4\times4$
and $8\times8$ luminance blocks. The position of the predicted coefficients
$\mathcal{I}_{DCT}$ is defined for each size of block and for each
intra prediction mode as illustrated on \green{Fig.} \ref{Fig:inpainted_coefficients_4x4}
and \green{Fig.} \ref{Fig:inpainted_coefficients_8x8}. For the descent gradient
algorithm used in the proposed method, we set the maximum number of
iterations to $100$, with a fixed regularization parameter. The implementation
of a more appropriate algorithm than the generic descent gradient
algorithm will be further investigated in order to converge toward
a solution in a faster way. The aim of this paper is primarily to
demonstrate that compression gains are possible with this method.
The Bjontegaard metric \cite{Bjontegaard2001} is used to measure
the gain against the H264/AVC reference software, with QP = 22, 27,
32, 37. The rate gains are expressed in percentage for a similar objective
quality.

The main advantage of our method is that no additional information
is coded, since the proposed method does not compete with another.
It's applied on every $4\times4$ and $8\times8$ intra blocks.

\begin{figure}[h]
\centering\includegraphics[width=1\linewidth]{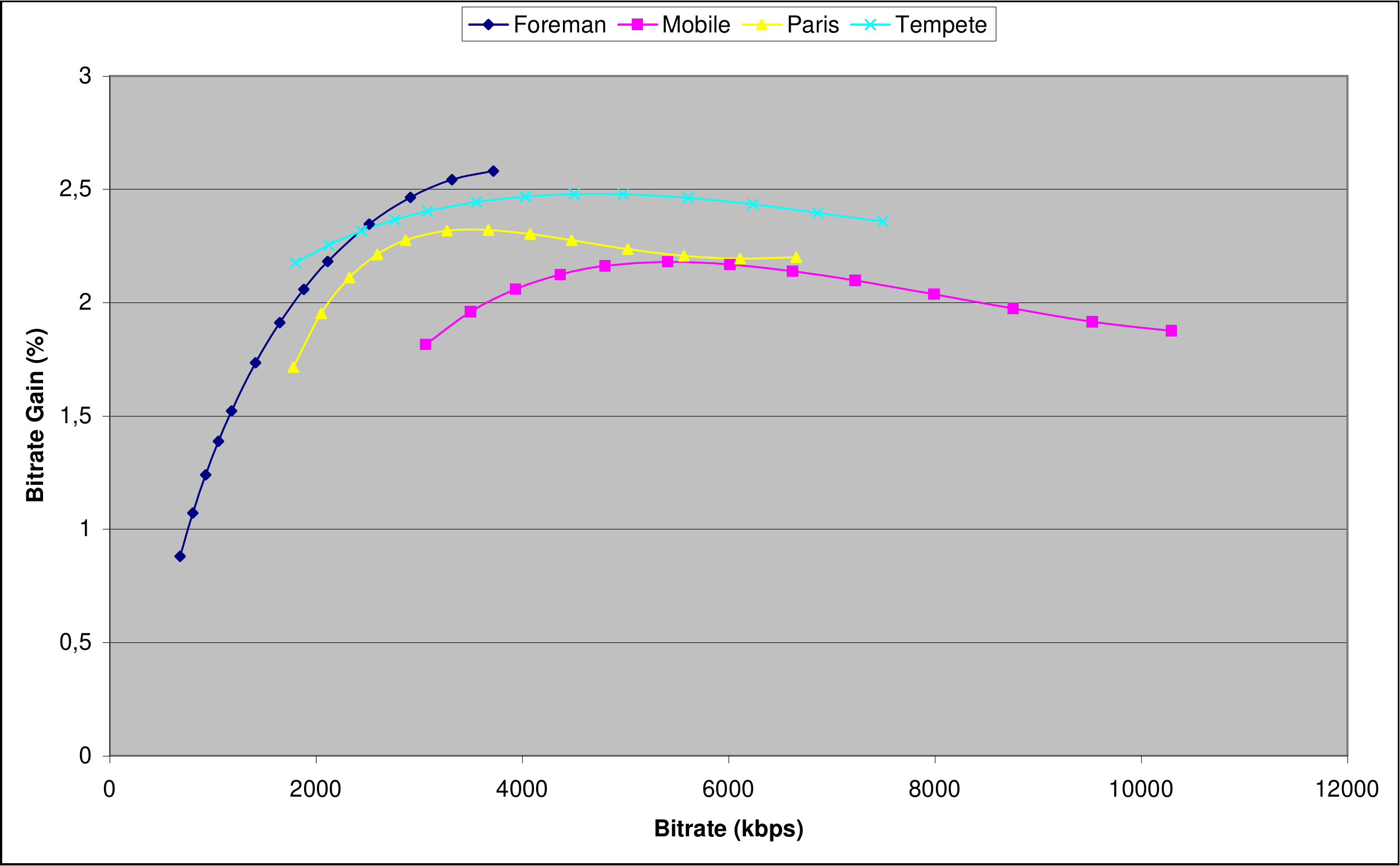}
\caption{Percentage of bitrate gain for four sequences according to the bitrate.}
\label{Fig:CIF_results}
\end{figure}

We have experimented our method on CIF resolution sequences ($352\times288$
pixels) and 720p sequences ($1280\times720$ pixels). The results
are presented on Tab. \ref{Tab:results}. In both CIF and 720p resolution,
the average bitrate gain is over $2\%$. In \green{Fig.}\ref{Fig:CIF_results},
we illustrate the bitrate gain according to the bitrate for the CIF
sequences. It is interesting to notice that at the extremum low and
high bitrates, the bitrate gains tend to decrease (see Foreman under
$1000kbps$). This is related to the fact that the set of predicted
coefficients $\mathcal{I}_{DCT}$ is certainly not adapted under those
conditions. For example, at very low bitrate, most of the coefficients
we want to predict are already set to zero, so no bitrate saving is
possible in this case. Worse, it may happen that the coefficient prediction
does not give a zero value, hence the prediction residue is more costly
to encode than the original non predicted coefficient.

\begin{algorithm}[h]\label{al:prediction-DCT}
\begin{enumerate}
\small{
\item Let $p$ be the intra or inter image predictor of the current block $u_{B}$
\item Let $r=DCT\left(u_{B}-p\right)$ be the transformed residual of $p$, and
$\widehat{r}$ the residual $r$ decoded for every coefficient $\vec{k}\in \mathcal{I}_{O}$
\item Let $L$ the maximum number of iterations
\item \textbf{While $\mathcal{I}_{DCT}\neq\emptyset$, Do}

\begin{enumerate}
\item Let $\beta_{\vec{k}}=\qquad \\ \left\{
\begin{array}{ccc}
\widehat{r}_{\kvec} \; \textrm{if} \;\vec{k}\in \mathcal{I}_{O} \; \textrm{(coeff. not to predict)}\\
0 \; \textrm{otherwise} \;\textrm{(coeff. to predict)}
\end{array}\right.$
\item Let $\vec{c}\in \mathcal{I}_{DCT}$ the index of the coefficient to predict, with
$\vec{c}$ corresponding to the first DCT coefficient of the block along the zigzag path
\item Let $n=0$ be the current iteration and $\gamma_{n}$ the gradient descent increment
for the prediction of the coefficient $\vec{\beta}$
\item \textbf{While $n<K$, Do}
\begin{enumerate}
\item Let $s= \qquad\\  DCT\left(curv\left(p+ DCT^{-1}\left(\beta^{n}\right)\right)\right)$
\item $\beta_{\vec{c}}^{n+1}=\beta_{\vec{c}}^{n}+\gamma_{n}s_{\vec{c}}$
\item $n=n+1$
\end{enumerate}
\item \textbf{End of Loop\label{enu:prediction-DCT-fin-boucle}}
\item Code/decode the residual of the intra-bloc prediction $r_{\vec{c}}-\beta_{\vec{c}}^{n+1}$
of the coefficient $\vec{c}$ to get $\widehat{r_{\vec{c}}-\beta_{\vec{c}}^{n+1}}$,
then reconstruct $\widehat{r_{\vec{c}}}=\beta_{\vec{c}}^{n+1}+\widehat{r_{\vec{c}}-\beta_{\vec{c}}^{n+1}}$
\item Update $\mathcal{I}_{DCT}$ and $\mathcal{I}_{O}$ so that $\mathcal{I}_{DCT}=\mathcal{I}_{DCT}/\vec{c}$,
$\mathcal{I}_{O}=\mathcal{I}_{O}\cup\vec{c}$
\end{enumerate}
\item \textbf{End of Loop}
} 
\end{enumerate}
\caption{Intra-block prediction method used in a video coding context.}
\end{algorithm}

\begin{table}
\centering%
\begin{tabular}{|l|c|}
\hline
sequence & Bitrate gain (\%)\tabularnewline
\hline
\hline
CIF & \tabularnewline
\hline
~~~Foreman & $1.88$\tabularnewline
\hline
~~~Mobile & $2.06$\tabularnewline
\hline
~~~Paris & $2.20$\tabularnewline
\hline
~~~Tempete & $2.40$\tabularnewline
\hline
AVERAGE CIF & $\mathbf{2.13}$\tabularnewline
\hline
\hline
720p & \tabularnewline
\hline
~~~BigShip & $1.40$\tabularnewline
\hline
~~~City & $1.77$\tabularnewline
\hline
~~~Raven & $2.02$\tabularnewline
\hline
~~~Night & $2.05$\tabularnewline
\hline
~~~ShuttleStart & $2.51$\tabularnewline
\hline
~~~OldTownCross & $\mathbf{2.58}$\tabularnewline
\hline
AVERAGE 720p & $\mathbf{2.05}$\tabularnewline
\hline
\end{tabular}
\caption{Percentage of bitrate gain according to the sequence, computed with
the Bjontegaard metric. Only $49$ first I frames (intra) are encoded.}
\label{Tab:results}
\end{table}

\begin{figure}[h]
\centering%
\begin{tabular}{cc}
\subcaptionbox{$QP=22$ (high-rate)}{\includegraphics[width=0.45\linewidth]{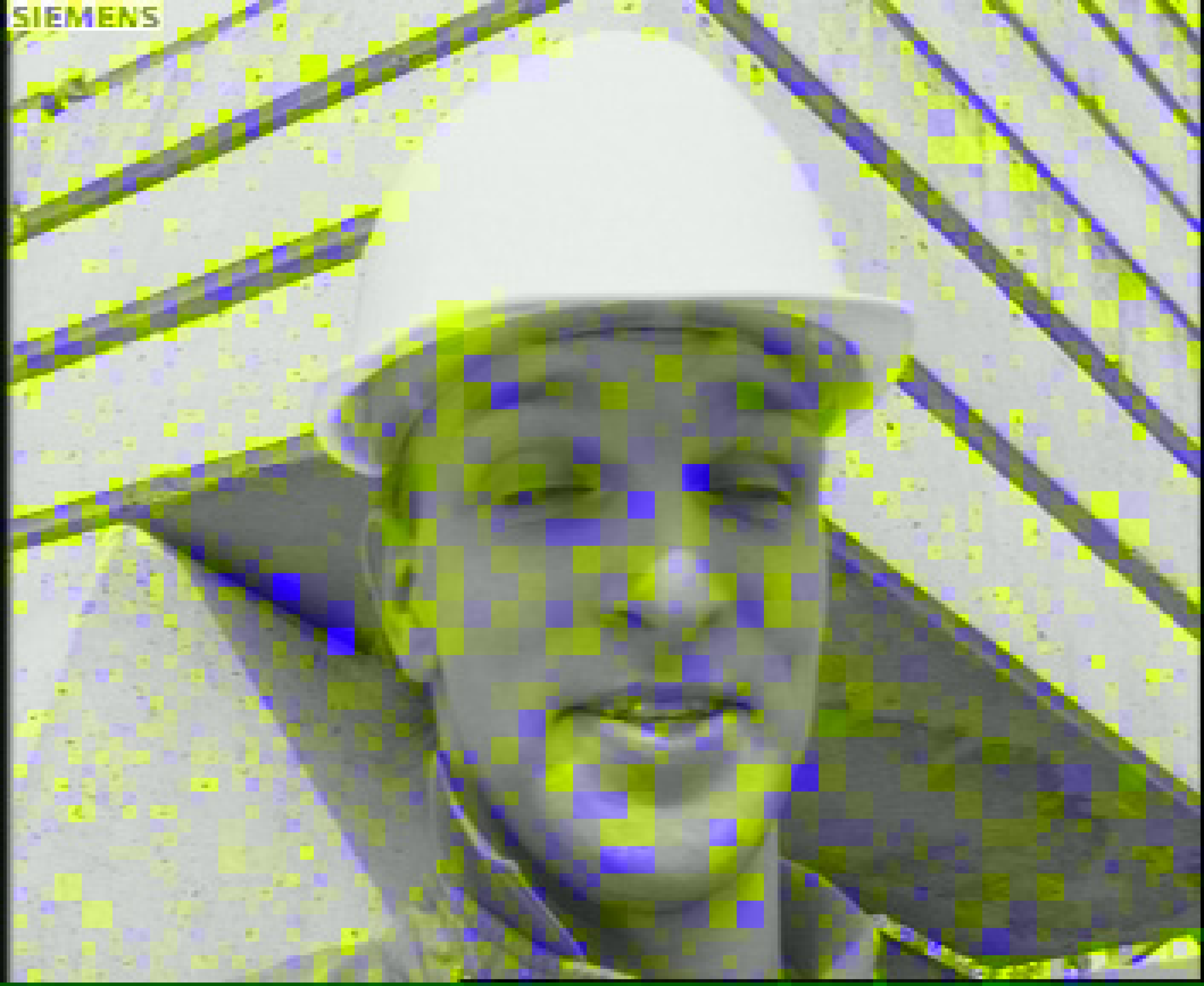}} & \subcaptionbox{$QP=27$}{\includegraphics[width=0.45\linewidth]{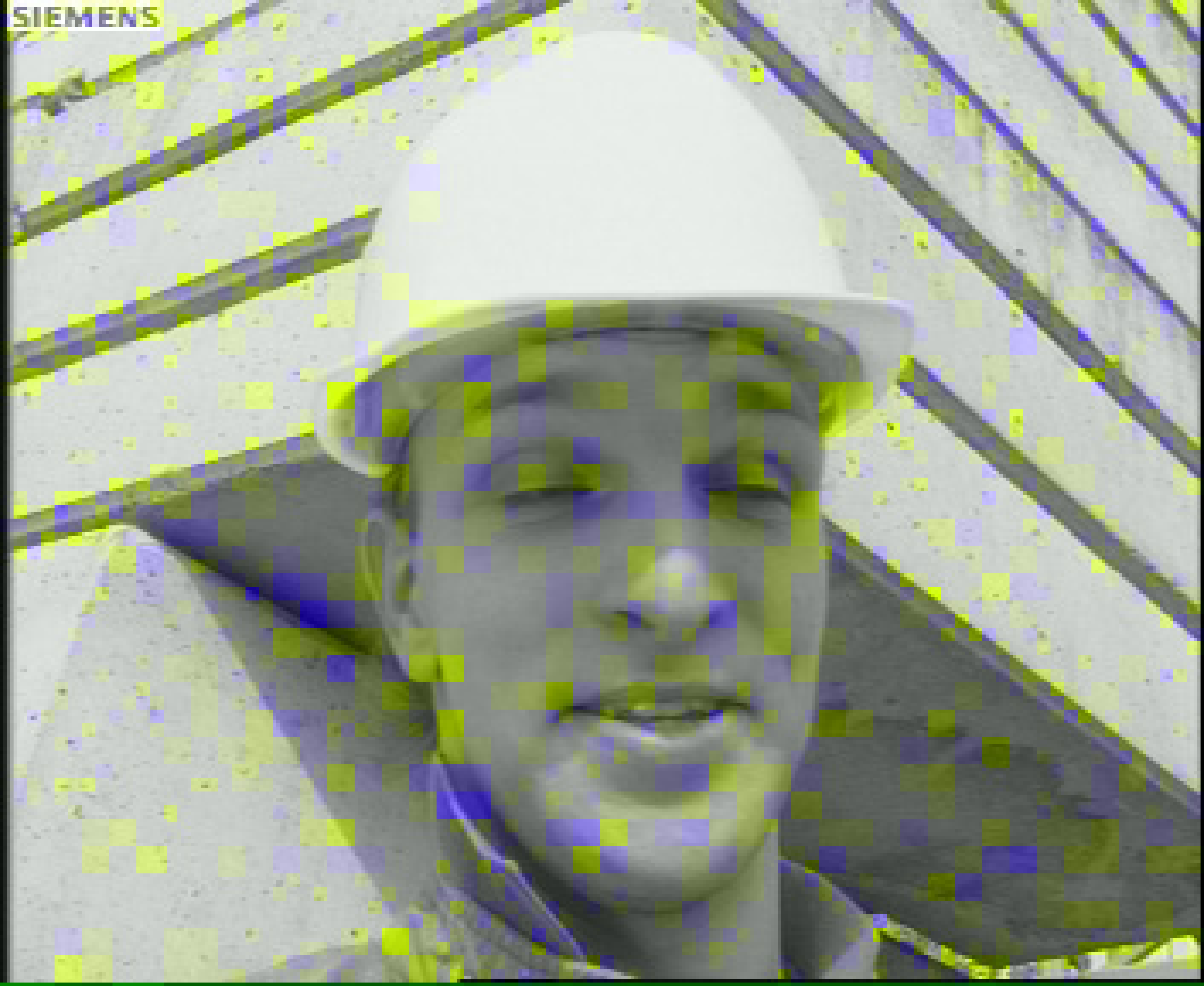}}\tabularnewline
\subcaptionbox{$QP=32$}{\includegraphics[width=0.45\linewidth]{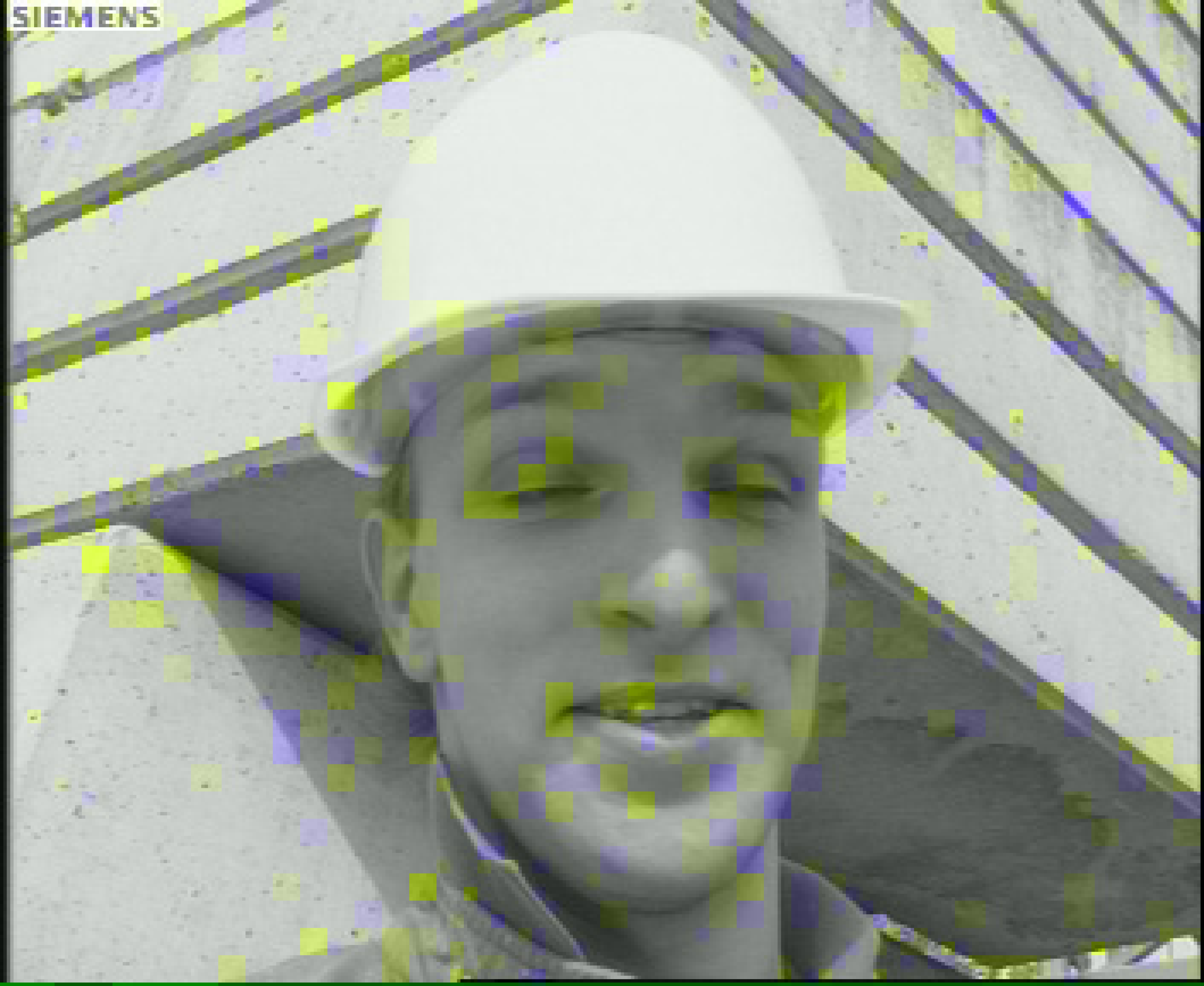}} & \subcaptionbox{$QP=37$(low-rate)}{\includegraphics[width=0.45\linewidth]{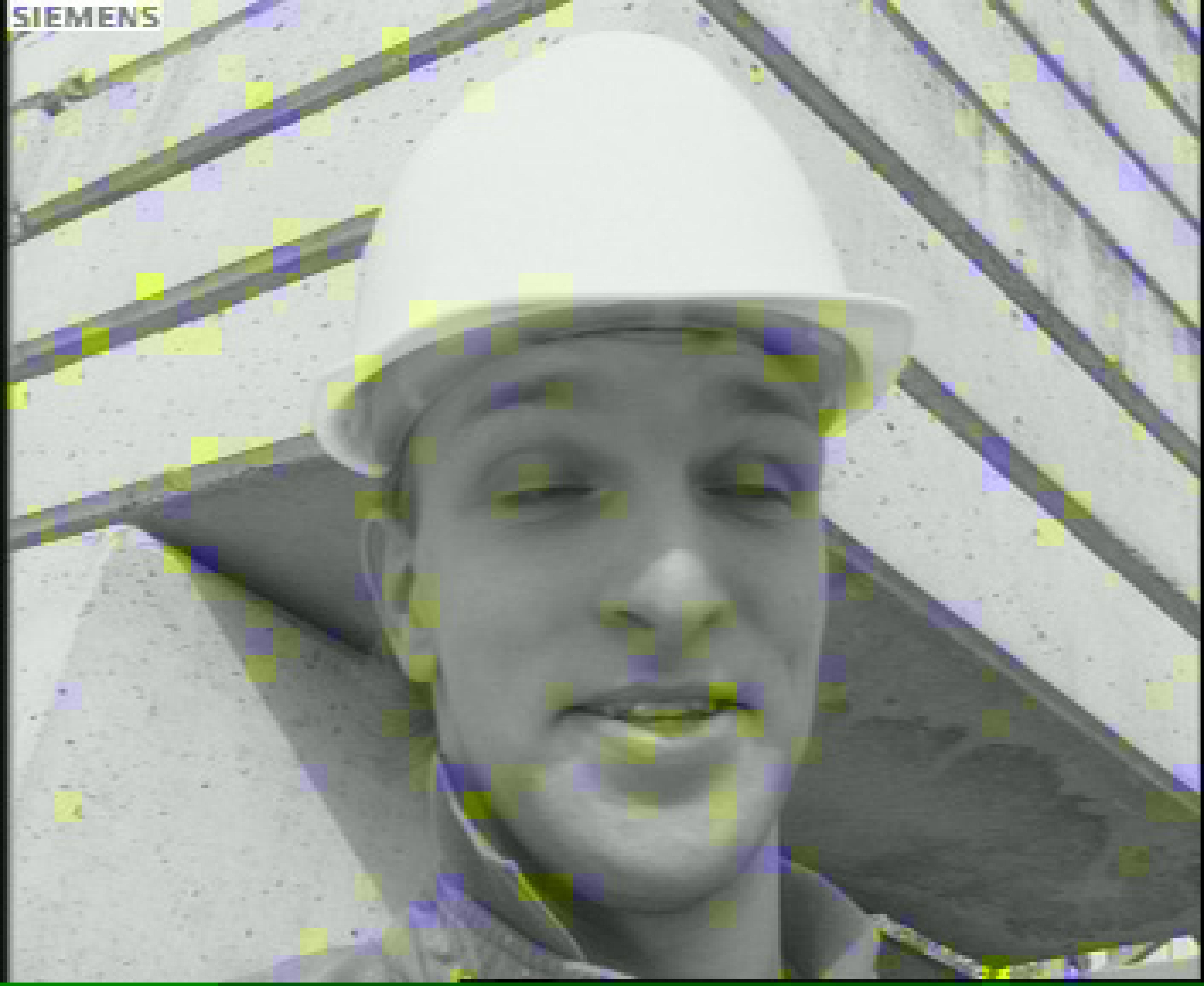}}\tabularnewline
\end{tabular}

\caption[Résultats obtenus sur la première image de la séquence \textit{Foreman}.]{Results on the first frame of the \textit{Foreman} sequence. In green~: the blocks where the intra-block prediction reduces the number of coded symbols (here w.r.t. H264/AVC). In blue~: when the number of coded symbols increases; the higher the chroma intensity, the higher the difference.}
\label{Fig:DCT-pred-foreman}
\end{figure}

\begin{figure}[h]
\centering
\begin{tabular}{cc}
\subcaptionbox{$QP=22$ (high-rate)}{\includegraphics[width=0.45\linewidth]{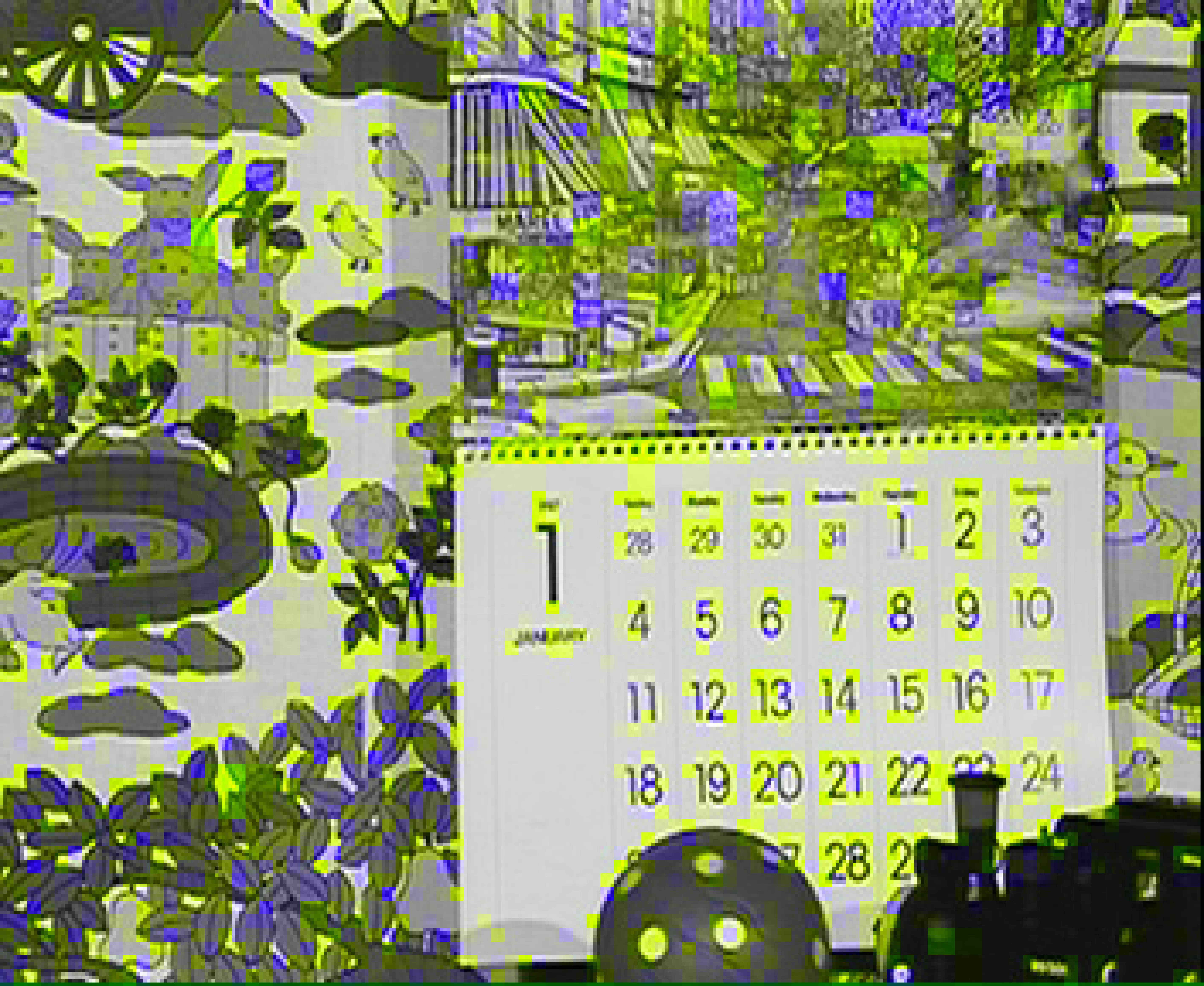}}
&\subcaptionbox{$QP=27$}{\includegraphics[width=0.45\linewidth]{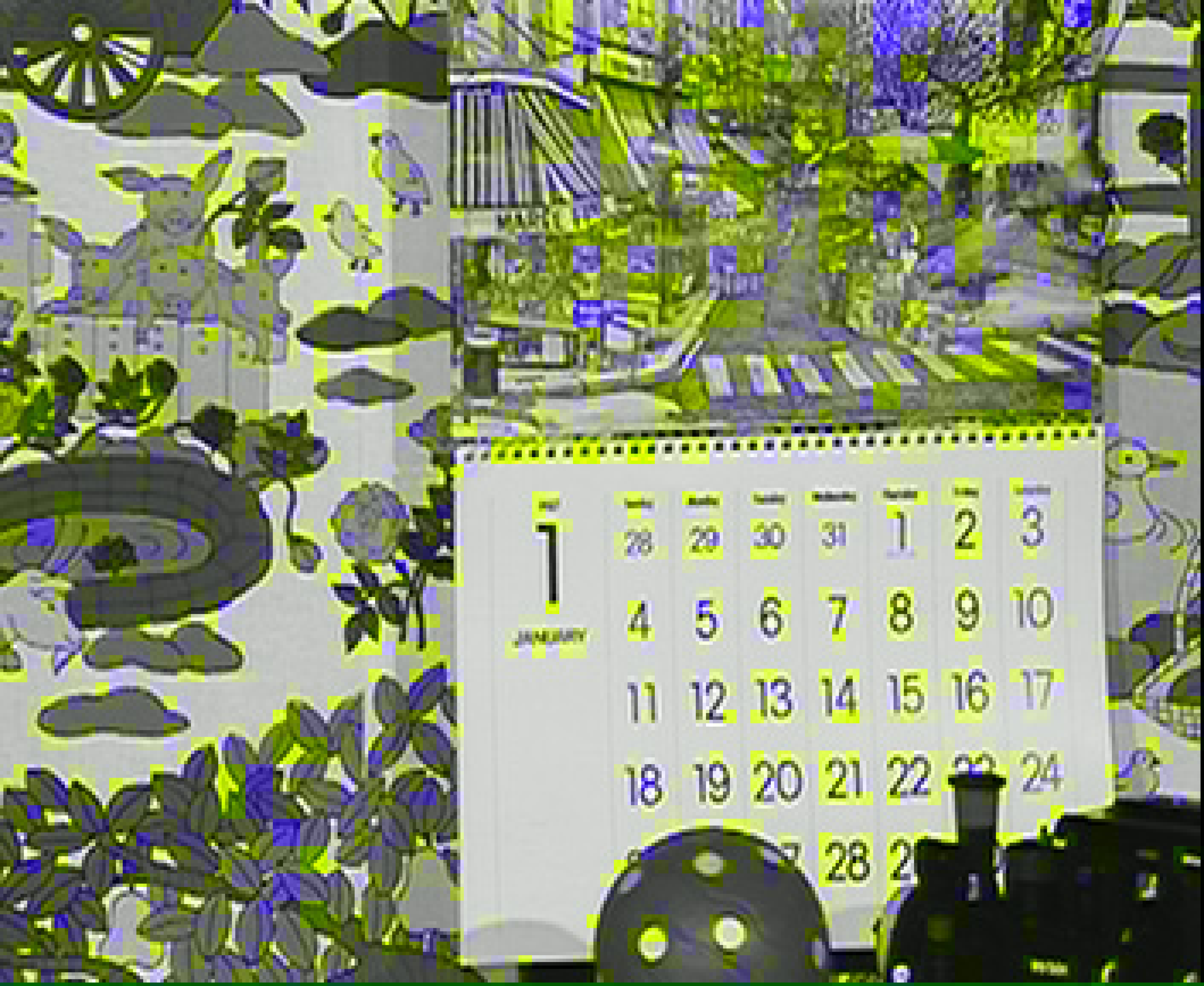}}\tabularnewline
\subcaptionbox{$QP=32$}{\includegraphics[width=0.45\linewidth]{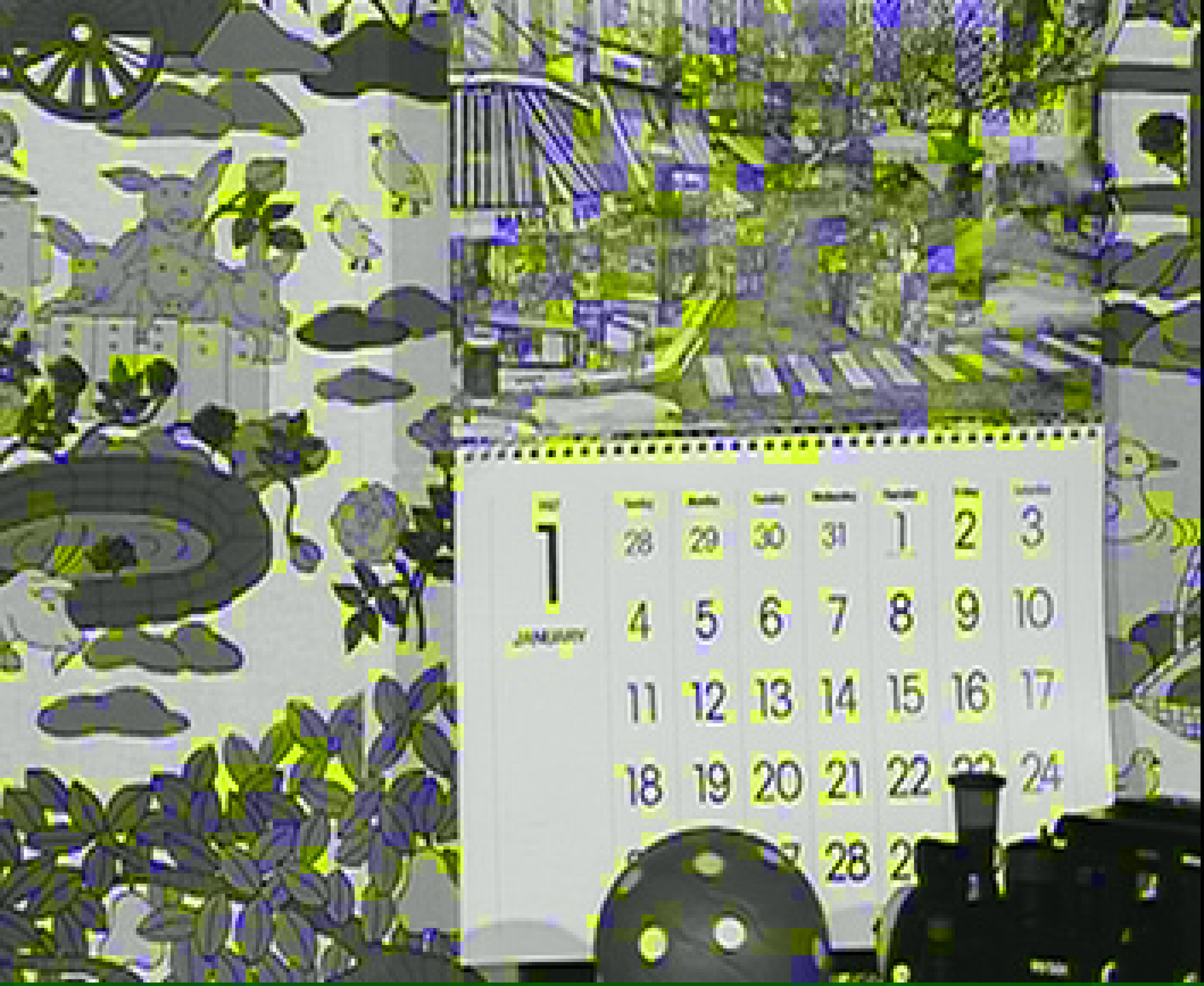}}
&\subcaptionbox{$QP=37$(low-rate)}{\includegraphics[width=0.45\linewidth]{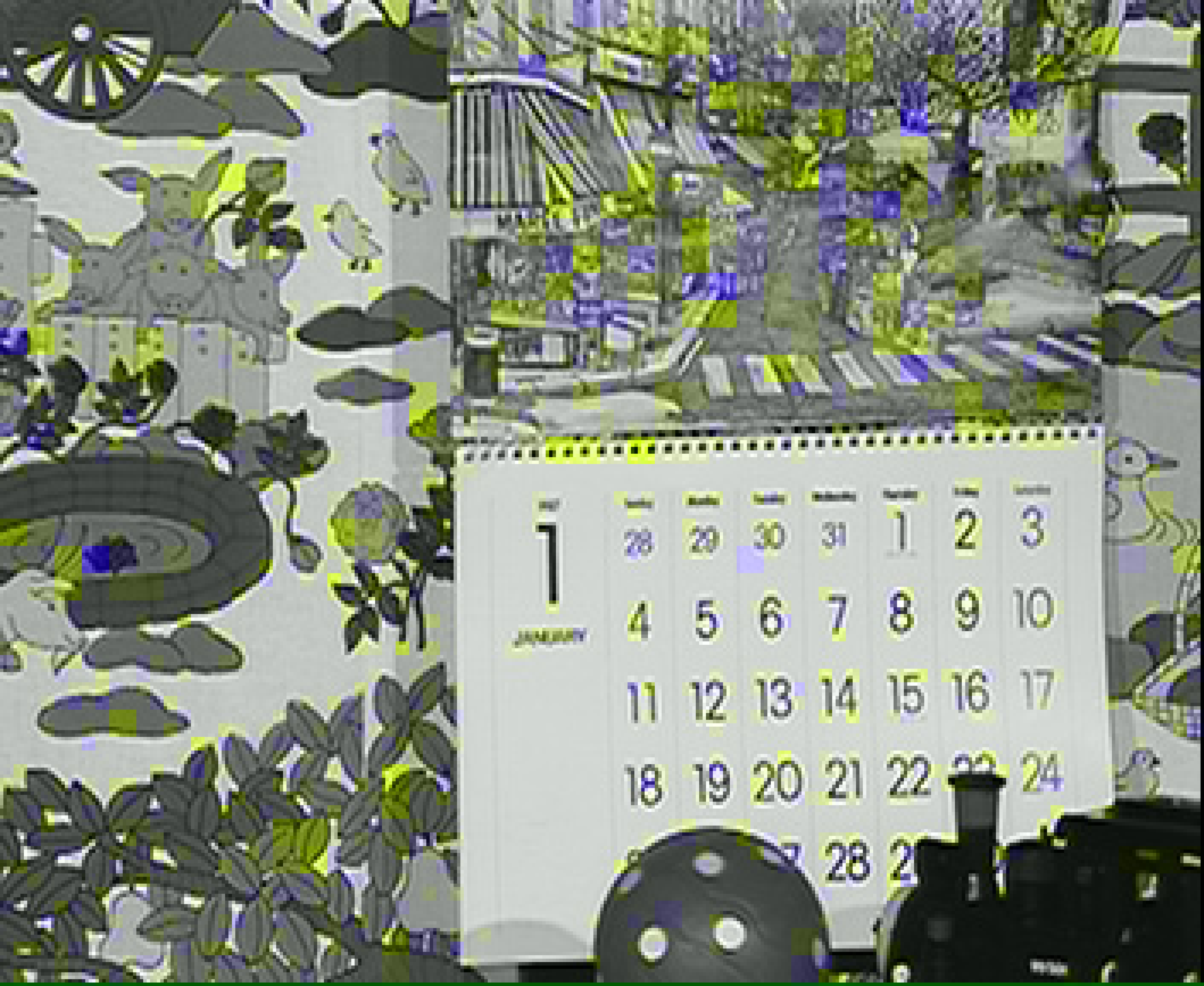}}\tabularnewline
\end{tabular}
\caption[Résultats obtenus sur la première image de la séquence \textit{Mobile}.]{Results obtained on the first image of the \textit{Mobile} (or \textit{caltrain}) sequence. In green~: the blocks where the intra-block prediction reduces the number of coded symbols. In blue~: when the number of coded symbols increases. The higher the chroma intensity, the higher the difference.}
\label{Fig:DCT-pred-mobile}
\end{figure}

\begin{figure*}[!t]
\includegraphics[width=1\linewidth]{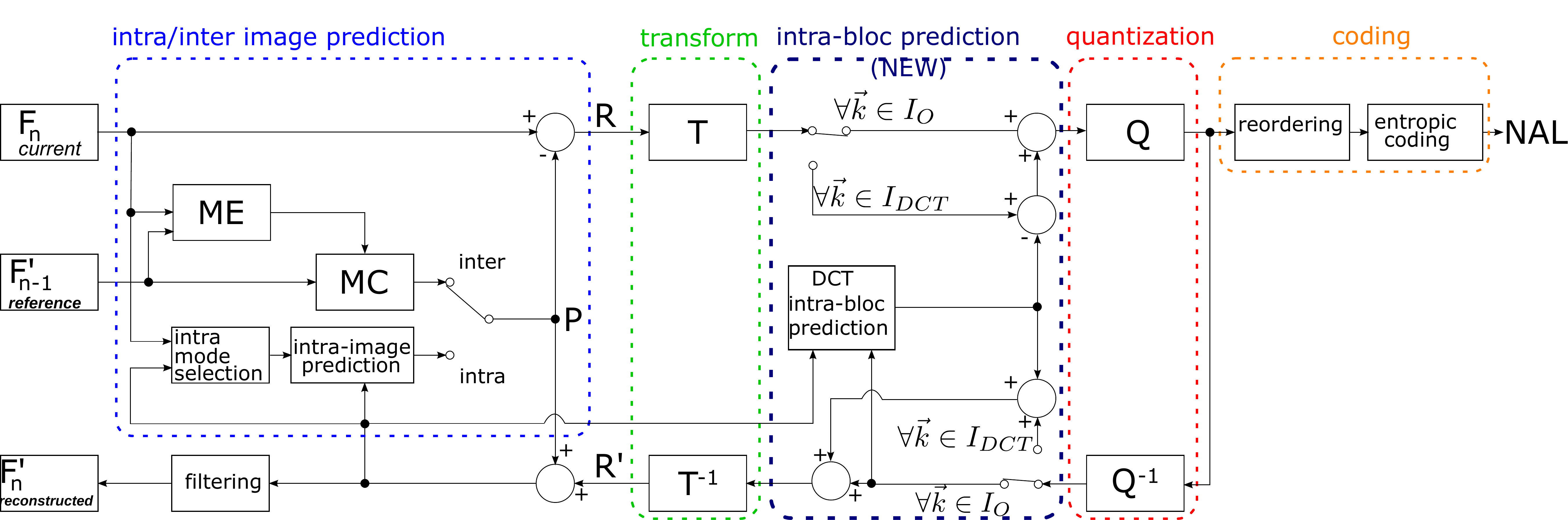}
\caption{Video encoding scheme, based on a hybrid block coding and a linear residual transform, and using the new intra-block prediction stage. }
\label{Fig:new_encoder_schema}
\end{figure*}

Finally the \green{Fig.} \ref{Fig:DCT-pred-foreman} shows the gains obtained per block on the first image of the \textit{Foreman} sequence and for different bit rates. We can remark 1) that the highest disparity in coding costs happens for high rates. The quantization step being lower in this case, there are more non null coefficients to encode, thus more coefficients on which our prediction method can bring interesting results. 2) For some blocks, our method can generate an extra information to encode, thus the intra-block prediction residue exhibits a higher amplitude than the original coefficient. However, it seems that some neighbor blocks, with very similar aspects, finally give very different results in terms of coding costs. We thus conclude that there is no ``typical'' region (texture, contour ...), for which our method presents a better yield, except for flat regions where no coding cost difference is observable. This is even more obvious with the Mobile image on
\green{Fig.} \ref{Fig:DCT-pred-mobile}.

\section{Conclusion}
We have investigated the domain of analysis/synthesis and, in particular of inpainting methods, to improve current video compression algorithms. To this end we have first developed a regularization model, based on total variation minimization, for the restoration of DCT coefficients in a Block-DCT (B-DCT) compressed image. This model has been efficiently tested for the restoration of images subject to quantization and block-based artifacts. We then investigated the compression domain by developing a selective cancellation/restoration scheme of DCT coefficients in a JPEG image. In this framework we noticed that a two, low-frequency, residual coefficients cancellation/restoration scheme is the optimal configuration for improving the compression. \\
Encouraging results on JPEG leaded us to integrate this method as a second, supplementary, prediction stage, to the standard spatial prediction stage of the H264/AVC block coder/decoder. We have modified the residual DCT coding stage to integrate our new ``Video Coding Residual Prediction'' (VcResPred) method. For the video coding scheme we do not use any more a two low-frequency static DCT coefficients predictive set~: our algorithm has been optimized for the coding standard considered by i) using the knowledge of the spatial prediction mode to adaptively select the set of predicted coefficients
.... ii) using an iterative scheme to optimally predict the coefficients of a same block iii) modifying the standard zig-zag path for optimally predict these coefficients.
A key result of i), i.e. using the spatial prediction mode to select the DCT coefficients to predict, is that it does not require any additional signalization towards the decoder, thus there is no additional payload in this compression scheme.
The experiments conducted with the integration of this scheme in H264/AVC have shown a significative reduction of the bitrate for the same PSNR~: the average \textit{objective} improvement in
bitrate savings is over $2\%$, with a maximum of $2.6\%$, a percentage which is a key result for the integration in a standard video codec. We thus fully reached our main objective of rate/distorsion gain.

%
As the complexity was not our main concern we have not looked at an optimized scheme until now. Nevertheless we must precise that several ways can be adopted to minimize this one~: first the number of DCT/IDCT can be lowered by processing only one coefficient at each iteration and not the whole block. Moreover, the number of iterations and gradient descent step are fixed. These two parameters could be adjusted and we could use a loop output constraint on a steady state.

As final remarks, we must first mention that the same scheme could as well be used for an inter-frame prediction, but with probably a less
significant gain per frame due to the weak energy embodied in inter-frame coefficients. The second, fundamental, point is that this
transform-domain prediction method could as well be implemented for other kinds of linear block-transforms than B-DCT, like the
``DST-style'' transform of HEVC.

The first perspective for this work is, obviously, to fully integrate our cancellation/restoration principle into the HEVC scheme. As second perspective, we aim to study more accurately the regularization model and in particular the total variation criterion which is supposed to be suboptimal for a large number of images. More sophisticated models of PDEs, like the one used in anisotropic diffusion \cite{Galic2008}, could provide better results, especially for the case of large CTBs (Coding Tree Blocks). Also a structure + texture decomposition (e.g. Meyer's model \cite{Meyer2001}) scheme, combined with texture synthesis algorithms (Bertalmio et al. \cite{Bertalmio2003}), will probably be investigated for a new inpainting model.
We end this conclusion by notifying that this work already led to the publication of an international patent \cite{Amonou2012}.

%


\bibliographystyle{plain}
\bibliography{refs_DCTpredict_pb_ArXiv}

\begin{thebibliography}{10}

\bibitem{Alter2004}
F.~Alter, S.~Durand, and J.~Froment.
\newblock Deblocking {DCT}-based compressed images with weighted total
  variation.
\newblock In {\em IEEE International Conference on Acoustics, Speech, and
  Signal Processing (ICASSP '04)}, volume~3, pages iii--221--4 vol.3, May 2004.

\bibitem{Alter2005}
F.~Alter, S.~Durand, and J.~Froment.
\newblock Adapted total variation for artifact free decompression of {JPEG}
  images.
\newblock {\em J. Math. Imaging Vis.}, 23(2):199--211, 2005.

\bibitem{Amonou2012}
I.~Amonou, M.~Moinard, P.~Duhamel, and P.~Brault.
\newblock International patent~: {Methods and Devices for Encoding and Decoding
  at least One Image Implementing an Estimation in the Transform Domain, and
  Corresponding Signal and Computer Program}.
\newblock PCT/FR2011/051383- WO/2012/001257, January 2012.

\bibitem{Bertalmio2003}
M.~Bertalmio, L.~Vese, G.~Sapiro, and S.~Osher.
\newblock Simultaneous structure and texture image inpainting.
\newblock {\em Computer Vision and Pattern Recognition, 2003. Proceedings. 2003
  IEEE Computer Society Conference on}, 2:II -- 707--12 vol.2, june 2003.

\bibitem{Bjontegaard2001}
G.~Bjontegaard.
\newblock Calculation of average {PSNR} differences between {RD} curves.
\newblock In {\em ITU-T SC16/Q6}, document VCEG-M33, USA, April 2001. 13th VCEG
  Meeting.

\bibitem{Bourdon2005}
P.~Bourdon, B.~Augereau, C.~Chatellier, and C.~Olivier.
\newblock Transmission and quantization error concealment for jpeg color images
  using geometry-driven diffusion.
\newblock In {\em IEEE International Conference on Image Processing, ICIP
  2005}, volume~1, pages I -- 801--4, sept. 2005.

\bibitem{Chambolle2004}
A.~Chambolle.
\newblock An algorithm for total variation minimization and applications.
\newblock {\em Journal of Mathematical Imaging and Vision}, 20(1):89--97,
  January 2004.

\bibitem{Chambolle1997}
A.~Chambolle and P.~L. Lions.
\newblock Image recovery via total variation minimization and related problems.
\newblock {\em Numerische Mathematik}, 76:167--188, 1997.

\bibitem{Chan1999}
T.-F. Chan, G.H. Golub, and P.~Mulet.
\newblock A nonlinear primal-dual method for total variation-based image
  restoration.
\newblock {\em SIAM Journal of Scientific Computing}, 20(6):1964--1977, 1999.

\bibitem{Chan2002}
T.~F. Chan, S.H. Kang, , and J.~Shen.
\newblock Euler's elastica and curvature based inpaintings.
\newblock {\em SIAM Journal Applied Math.}, 63:564--592, 2002.

\bibitem{Chan2001a}
T.F. Chan, S.J. Osher, and J.~Shen.
\newblock The digital tv filter and nonlinear denoising.
\newblock {\em Image Processing}, 10(2):231--241, February 2001.

\bibitem{Chan2001}
T.F. Chan and J.~Shen.
\newblock Non-texture inpaintings by curvature-driven diffusions.
\newblock {\em J. Visual Comm. Image Rep.}, 12(4):436--449, 2001.

\bibitem{Chan2002b}
T.F. Chan and J.~Shen.
\newblock Mathematical models for local non-texture inpaintings.
\newblock {\em SIAM Journal of Applied Math.}, 63(3):1019--1043, 2002.

\bibitem{Chan2006}
T.F. Chan, J.~Shen, and H.M. Zhou.
\newblock Total variation wavelet inpainting.
\newblock {\em J. Math. Imaging Vision}, 25:107--125, 2006.

\bibitem{Chan1998}
T.F. Chan and C.K. Wong.
\newblock Total variation blind deconvolution.
\newblock {\em Image Processing}, 7(3):370--375, March 1998.

\bibitem{Chan2000}
T.F. Chan and H.M. Zhou.
\newblock Optimal constructions of wavelet coefficients using total variation
  regularization in image compression.
\newblock In {\em Image Compression, CAM Report, No. 00-27, Dept. of Math.,
  UCLA}, 2000.

\bibitem{Galic2008}
I.~Galic, J.~Weickert, M.~Welk, A.~Bruhn, A.~Belyaev, and H.-P. Seidel.
\newblock Image compression with anisotropic diffusion.
\newblock {\em {Journal of Mathematical Imaging and Vision}, Vol. 31, 255–269,
  2008}, 31(2-3):255–269, 2008.

\bibitem{Goldstein2009}
T.~Goldstein and S.~Osher.
\newblock The split-{B}regman method for {L1}-regularized problems.
\newblock {\em SIAM Journal of Image Science}, 2(2):323--343, 2009.

\bibitem{ITU-T_JCTVC}
{ITU-T SG16 WP3 and ISO/IEC JTC1/SC29/WG11}.
\newblock ``{Test Model under Consideration}'', {ITU-T/ISO/IEC} joint
  collaborative team on video coding ({JCT-VC}).
\newblock JCTVC-A205, April 2010.

\bibitem{Meyer2001}
Y.~Meyer.
\newblock {\em Oscillating Patterns in Image Processing and Nonlinear Evolution
  Equations: The Fifteenth Dean Jacqueline B. Lewis Memorial Lectures}.
\newblock American Mathematical Society, Boston, MA, USA, 2001.

\bibitem{Moinard011}
M.~Moinard, I.~Amonou, P.~Brault, and P.~Duhamel.
\newblock Image and video compression scheme based on the prediction of
  transformed coefficients.
\newblock In {\em {IEEE} International Conference on Signal Processing and
  Applications, {ISPA011}}, pages 385--389, {4-6} September 2011.

\bibitem{Park1997}
J.W. Park, J.W. Kim, and S.U. Lee.
\newblock {DCT} coefficients recovery based error concealment technique and its
  application to the {MPEG-2} bit stream error.
\newblock {\em IEEE Transactions on Circuits and Systems for Video Technology},
  7(6):845--854, December 1997.

\bibitem{Pennebaker1992}
W.B. Pennebaker and J.L. Mitchell.
\newblock {\em JPEG Still Image Data Compression Standard}.
\newblock Kluwer Academic Publishers, Norwell, MA, USA, 1992.

\bibitem{Richardson}
Iain~E. Richardson.
\newblock {\em {H.264} and {MPEG-4} Video Compression: Video Coding for Next
  Generation Multimedia}.
\newblock Wiley, 1st edition, August 2003.

\bibitem{Rudin1992}
L.I. Rudin, S.~Osher, and E.~Fatemi.
\newblock Nonlinear total variation based noise removal algorithms.
\newblock {\em Phys. D}, 60(1-4):259--268, 1992.

\bibitem{Rudin1994}
L.I. Rudin and S.J. Osher.
\newblock Total variation based image restoration with free local constraints.
\newblock In {\em ICIP}, pages I: 31--35, 1994.

\bibitem{Salama1998}
P.~Salama, N.B. Shroff, and E.J. Delp.
\newblock {\em Image Recovery Techniques for Image Compression Applications~:
  7. {Error} Concealment in Encoded Video Streams}.
\newblock Katsaggelos, Aggelos; Galatsanos, Nick (Eds.), Kluwer, 1st edition,
  1998.

\bibitem{Schwarz2005d}
H.~Schwarz, T.~Hinz, and K.~Suehring.
\newblock {Joint Scalable Video Model 9.7 }{(JSVM 9.7)}.
\newblock Technical report, Joint Video Team, 2005.

\bibitem{Schwarz2007}
H.~Schwarz, D.~Marpe, and T.~Wiegand.
\newblock Overview of the scalable extension of the {H.264/MPEG-4 AVC} video
  coding standard.
\newblock {\em IEEE Transactions on Circuits and Systems for Video Technology},
  2007.

\bibitem{Sullivan2012}
G.~J. Sullivan, J.-R. Ohm, W.-J. Han, and T.~Wiegand.
\newblock {Overview of the High Efficiency Video Coding}.
\newblock {\em {IEEE Trans. on Circuits and Systems}}, 22(12):1649--1668, 2012.

\bibitem{Vogel1996}
C.~R. Vogel and M.~E. Oman.
\newblock Iterative methods for total variation denoising.
\newblock {\em SIAM Journal of Scientific Computing}, 17(1):227--238, 1996.

\bibitem{Wang1991}
Y.~Wang and Q.F. Zhu.
\newblock Signal loss recovery in {DCT}-based image and video codecs.
\newblock {\em Visual Communications and Image Processing '91: Visual
  Communication}, 1605(1):667--678, 1991.

\bibitem{Wang1993}
Y.~Wang, Q.F. Zhu, and L.~Shaw.
\newblock Maximally smooth image recovery in transform coding.
\newblock {\em IEEE transactions on communications}, 41(10):1544--1551, 1993.

\bibitem{Wu2007}
X.~Wu, Q.~Sun, K.~Zhang, and L.~Yu.
\newblock Modeling natural image for estimating {DCT} coefficient properties of
  intra prediction.
\newblock In {\em ICME}, pages 476--479, 2007.

\bibitem{Zhong1996}
S.~Zhong.
\newblock Image compression by optimal reconstruction.
\newblock United States Patent 5,534,925, July 1996.
\newblock Cognitech Inc.

\end{thebibliography}
%


\end{document}